\documentclass[12pt]{article}
\usepackage[english,german]{babel}

\usepackage[inner=3.50cm,outer=3.50cm,top=2.75cm,bottom=2.75cm,includeheadfoot]{geometry}

\usepackage{lscape}
\usepackage{amsmath}
\usepackage{bbm}
\usepackage{amssymb}
\usepackage{xcolor}
\usepackage{appendix}
\usepackage{rotating}
\usepackage[percent]{overpic}
\usepackage{longtable}
\usepackage{sidecap}
\usepackage{theorem}
\usepackage{bm}
\usepackage{overpic}
\usepackage{upgreek}
\usepackage{morefloats}
\usepackage{capt-of}

\begin{document}

\allowdisplaybreaks

\selectlanguage{german}

%\markboth{Yannick Wunderlich}
%{Complete Experiments in Meson Photoproduction}

%\catchline{}{}{}{}{}
%
%%%%%%%%%%%%%%%%%%%%%%%%%%%%%%%%%%%%%%%%%%%%%%%%%%%%%%%%%%%%%%%%%%%%

\title{Mathematical aspects of phase rotation ambiguities in partial wave analyses%\footnote{For the title, try not to use more than 
%3 lines. Typeset the title in 10 pt roman, uppercase and 
%boldface.}
}

\author{Yannick Wunderlich \\ Helmholtz-Institut f"ur Strahlen- und Kernphysik,\\ Universit"at Bonn \\ Nussallee 14-16,
 53115 Bonn, Germany % \\ wunderlich@hiskp.uni-bonn.de % for the CBELSA/TAPS collaboration%\footnote{
%Typeset names in 8 pt roman, uppercase. Use the footnote to indicate the
%present or permanent address of the author.}
}

\date{September 7, 2017}

% \address{Helmholtz-Institut f"ur Strahlen- und Kernphysik, Universit"at Bonn\\ Nussallee 14-16,
% 53115 Bonn, %Nordrhein Westfalen,
% Germany%\footnote{State completely without abbreviations, the
% % affiliation and mailing address, including country. Typeset in 8 pt
% % italic.
% \\
% wunderlich@hiskp.uni-bonn.de}

%\author{SECOND AUTHOR}

%\address{Group, Laboratory, Address\\
%City, State ZIP/Zone, Country\\
%second\_author@domain\_name}

\maketitle

%\begin{history}
%\received{Day Month Year}
%\revised{Day Month Year}
%\end{history}

\selectlanguage{english}

\begin{abstract}

The observables in a single-channel $2$-body scattering 
problem remain invariant once the amplitude is multiplied by an overall 
energy- and angle-dependent phase. This invariance is known as the con\-ti\-nu\-um ambiguity. Also, 
mostly in truncated partial wave analyses (TPWAs), discrete ambiguities originating from complex conjugation 
of roots are known to occur. In this note, it is shown that the general
con\-ti\-nu\-um ambiguity mixes partial waves and that for 
scalar particles, discrete ambiguities are just a subset of con\-ti\-nu\-um 
ambiguities with a specific phase. A numerical method is outlined briefly, 
which can determine the relevant connecting phases.

% The Complete Experiment problem for photoproduction of single pseudoscalar mesons is reviewed briefly. It is argued that, once amplitudes are obtained from a Complete Experiment, an undetermined overall phase denies access to partial waves. Then, the Complete Experiment problem is discussed in the context of a truncated partial wave analysis.
%\keywords{Complete Experiment; polarization observables; photoproduction; single energy analysis.}
\end{abstract}

%\ccode{PACS numbers: 11.25.Hf, 123.1K}

%\setlength{\textwidth}{16cm}
%\setlength{\textheight}{20cm}
%\pagestyle{plain}
\pagenumbering{arabic}

\section{Introduction} \label{sec:Introduction}

We assume the well-known partial wave decomposition of the amplitude $A(W,\theta)$ for a $2 \rightarrow 2$-scattering process of spinless particles
\begin{equation}
 A \left(W, \theta \right) = \sum_{\ell = 0}^{\infty} (2 \ell + 1) A_{\ell} (W) P_{\ell} (\cos \theta) \mathrm{.} \label{eq:BasicInfinitePWExpansion}
\end{equation}
\clearpage

\begin{figure}[ht]
\centering
\includegraphics[width=0.30\textwidth]{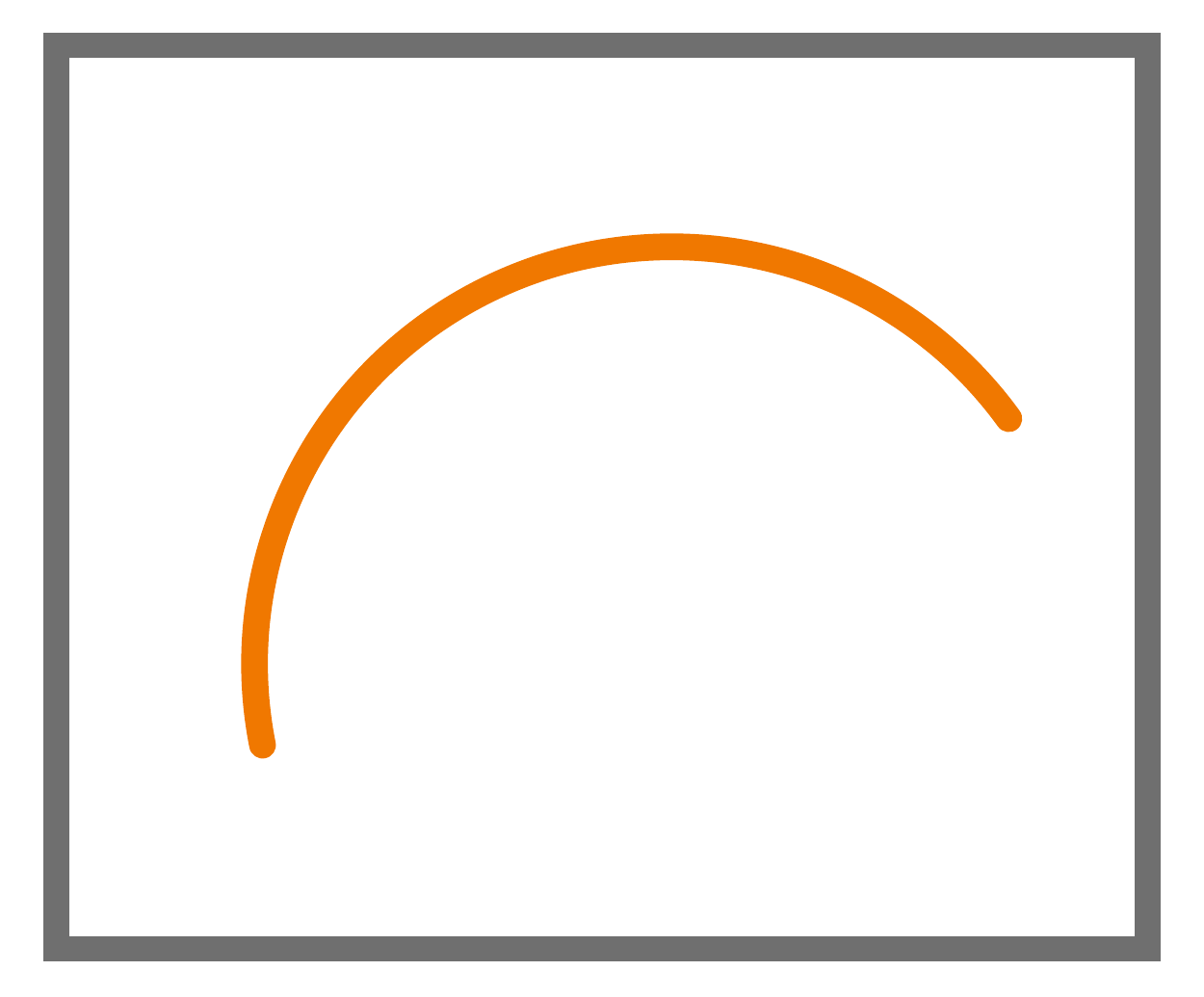}
\includegraphics[width=0.30\textwidth]{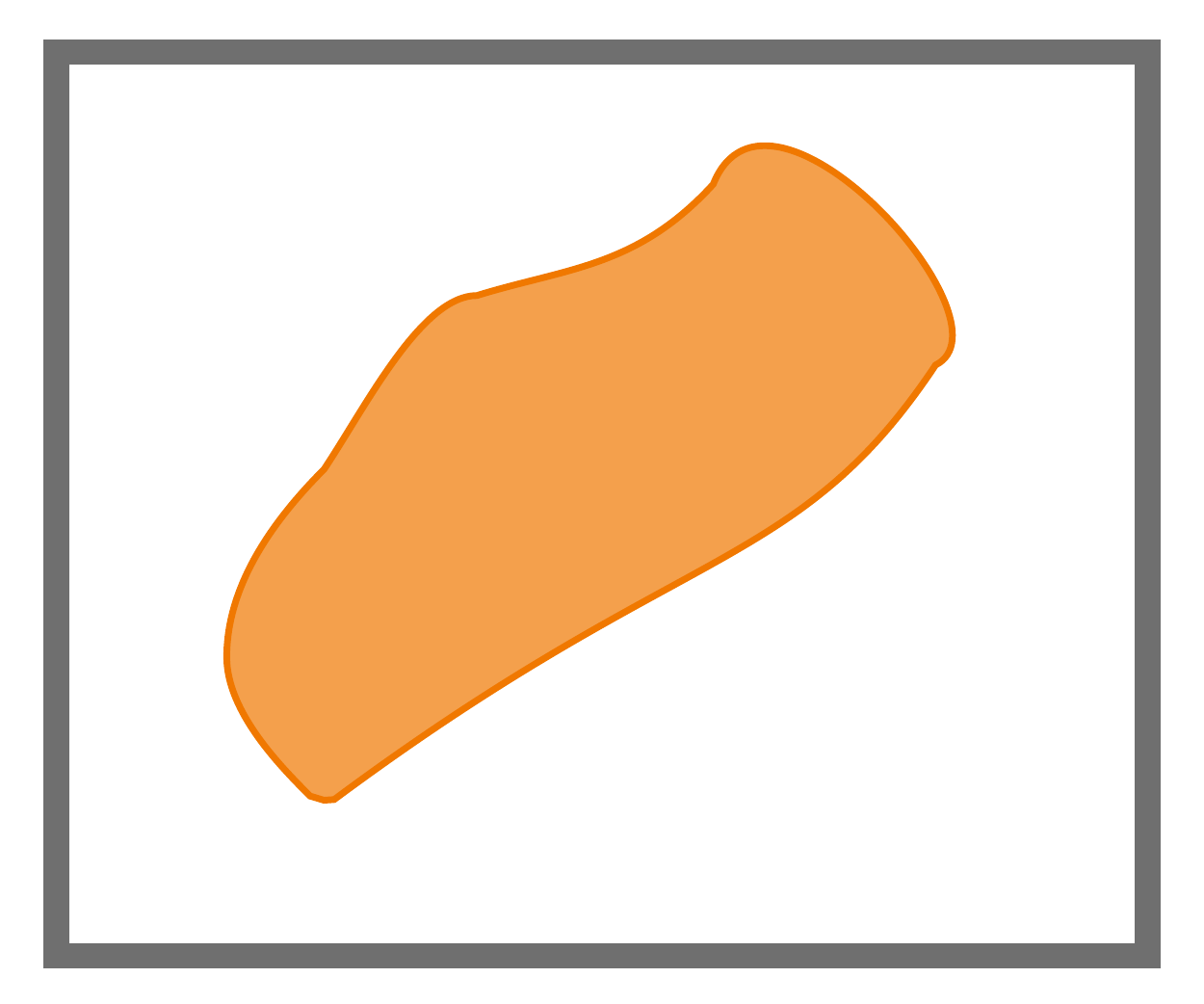}
\includegraphics[width=0.30\textwidth]{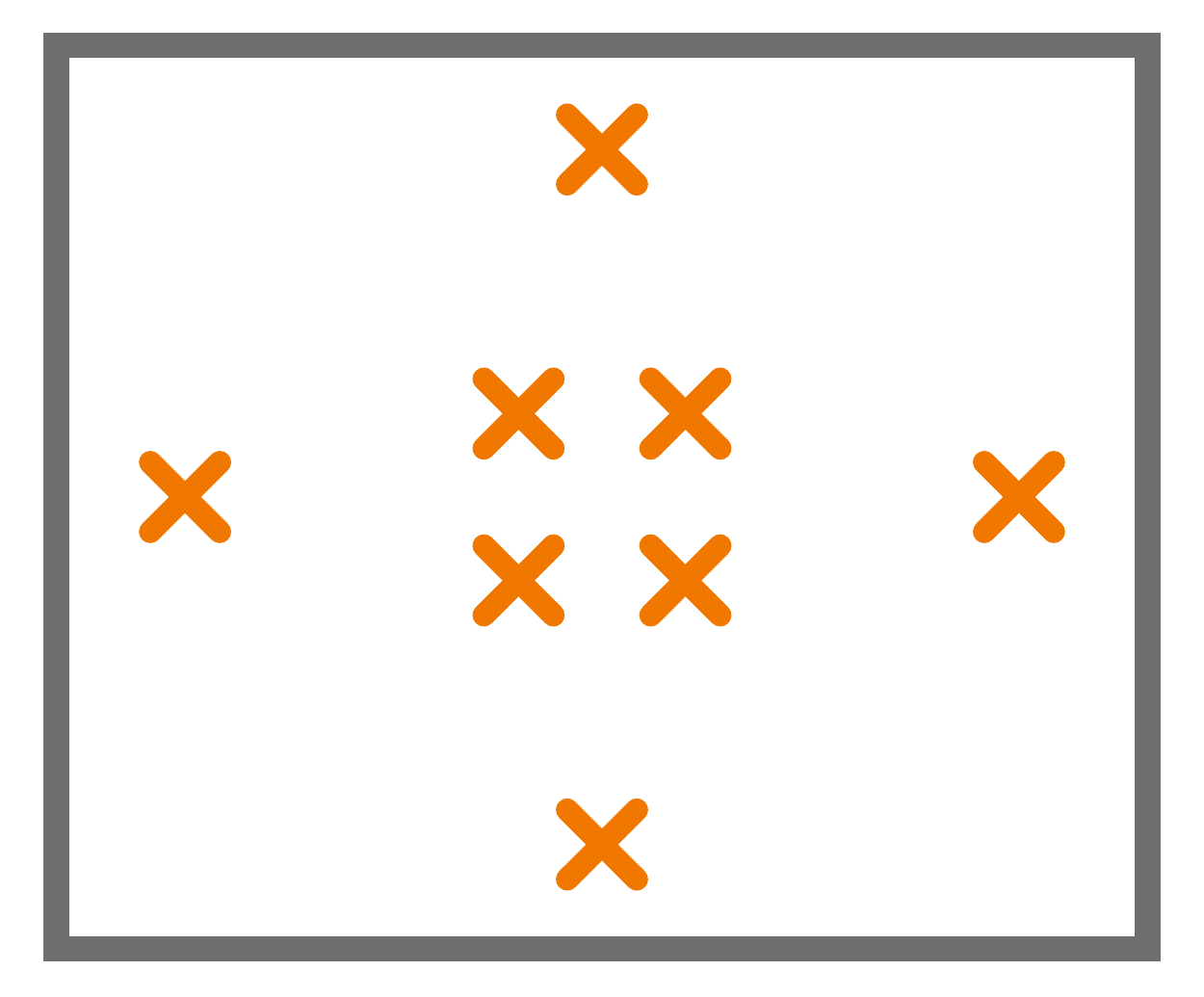}
\caption{Three schematic pictures are shown in order to distinguish the terms {\it discrete-} and {\it con\-ti\-nu\-um} ambiguities. The grey colored box depicts in each case the higher-dimensional parameter-space composed by the partial wave amplitudes, be it for infinite partial wave models, or for truncated ones. \newline
{\it Left:} One-dimensional (for instance circular) arcs can be traced out by con\-ti\-nu\-um ambiguity transformations, both for infinite and truncated models. \newline
{\it Center:} Connected continua in amplitude space, containing an infinite number of points with identical cross section, can be generated by use of angle-dependent rotations (\ref{eq:ContAmbTrafoDefinition}) (however, only in case the partial wave series goes to infinity). The connected patches are also called {\it islands of ambiguity} \cite{BowcockBurkhardt, AtkinsonEtAlContAmb}. \newline
{\it Right:} Discrete ambiguities refer to cases where the cross section is the same for discretely located points in amplitude space. These ambiguities are most prominent in TPWAs \cite{BowcockBurkhardt, Gersten}. However, two-fold discrete ambiguities can also appear for infinite partial wave models, once elastic unitarity is valid \cite{BowcockBurkhardt}. \newline
These figures have been published in reference \cite{Wunderlich2017}.
}
\label{fig:AmbiguityConceptSchematics}
\end{figure}

The {\it data} out of which partial waves shall be extracted are given by the differential cross section, which is (ignoring phase-space factors)
\begin{equation}
 \sigma_{0} \left(W, \theta \right) = \left| A \left(W, \theta \right) \right|^{2} \mathrm{.} \label{eq:DiffCSDefinition}
\end{equation}
Making a {\it complete experiment analysis} \cite{RWorkmanEtAlCompExPhotoprod} for this simple example, we see that the cross section constrains the amplitude to a circle for each energy and angle: $\left| A(W,\theta) \right| = + \sqrt{\sigma_{0} (W,\theta)}$. Thus, one energy- and angle-dependent phase is in principle unknown when based on data alone. The other side of the medal in this case is given by the fact that the amplitude itself can be rotated by an arbitrary energy- and angle-dependent phase and the cross section does not change. This invariance is known as the {\it con\-ti\-nu\-um ambiguity} \cite{BowcockBurkhardt}:
\begin{equation}
 A(W,\theta) \rightarrow \tilde{A} (W,\theta) := e^{i \Phi(W,\theta)} A(W,\theta) \mathrm{.} \label{eq:ContAmbTrafoDefinition}
\end{equation}
Another concept known in the literature on partial wave analyses is that of so-called {\it discrete ambiguities} \cite{BowcockBurkhardt, LPKokNote, Gersten}.
Suppose the full amplitude $A(W,\theta)$ can be split into a product of a linear-factor of the angular variable, for instance $\cos \theta$, and a remainder-amplitude $\hat{A} (W,\theta)$ \cite{LPKokNote}:
\begin{equation}
 A (W, \theta) =  \hat{A} (W,\theta) \left( \cos \theta - \alpha \right) \mathrm{.} \label{eq:DiscreteAmbKoksAmplitudeDecomposition}
\end{equation}
This is generally the case whenever the amplitude is a polynomial (i.e. the series (\ref{eq:BasicInfinitePWExpansion}) is truncated), but it may also be possible for infinite partial wave models. Then, it is seen quickly that the cross section (\ref{eq:DiffCSDefinition}) is invariant under complex conjugation of the root $\alpha$, which causes the discrete ambiguity
\begin{equation}
 \alpha \longrightarrow \alpha^{\ast} \mathrm{.} \label{eq:RootConjugationTrafo}
\end{equation}
Figure \ref{fig:AmbiguityConceptSchematics} shows a schematic illustration of the meaning of the terms {\it con\-ti\-nu\-um-} vs. {\it discrete ambiguities}. In this proceeding, the purely mathematical mechanisms (\ref{eq:ContAmbTrafoDefinition}) and (\ref{eq:RootConjugationTrafo}) are investigated. Of course, constraints from physics may reduce the amount of ambiguity encountered. For instance, unitarity is a very powerful constraint which, for elastic scatterings, leaves only one remaining non-trivial so-called {\it Crichton}-ambiguity \cite{Crichton}. This is believed to be true independent of any truncation-order $L$ of the partial wave expansion \cite{BowcockBurkhardt}. However, in energy-regimes where the scattering becomes inelastic, so-called {\it islands of ambiguity} are known to exist \cite{AtkinsonEtAlContAmb}. \newline
Although here we focus just on the scalar example, ambiguities have become a topic of interest in the quest for so-called {\it complete experiments} in reactions with spin, for instance photoproduction of pseudoscalar mesons \cite{RWorkmanEtAlCompExPhotoprod, YWEtAl2014}.

\clearpage

This proceeding is a briefer version of the more detailed publication \cite{Wunderlich2017}. The arXiv-reference \cite{Svarc2017} also treats very similar issues, as does the 
contribution of Alfred \v{S}varc to these Bled-proceedings.

%\clearpage

\section{The effect of con\-ti\-nu\-um ambiguity transformations on partial wave decompositions} \label{sec:GenEffectContAmb}

We let the general transformation (\ref{eq:ContAmbTrafoDefinition}) act on $A(W,\theta)$ and assume a partial wave decomposition for the original as well as the rotated amplitude
\begin{align}
 A(W,\theta) \longrightarrow \tilde{A} (W,\theta) &= e^{i \Phi(W,\theta)} A(W,\theta) = e^{i \Phi(W,\theta)} \sum_{\ell = 0}^{\infty} (2 \ell + 1) A_{\ell} (W) P_{\ell} (\cos \theta) \nonumber \\
 &\equiv \sum_{\ell = 0}^{\infty} (2 \ell + 1) \tilde{A}_{\ell} (W) P_{\ell} (\cos \theta) \mathrm{.} \label{eq:ContAmbTrafoPW}
\end{align}

\clearpage

Out of the infinitely many possibilities to parametrize the angular dependence of the phase-rotation, the convenient choice of a Legendre-series is employed
\begin{equation}
 e^{i \Phi (W, \theta)} = \sum_{k = 0}^{\infty} L_{k} (W) P_{k} (\cos \theta) \mathrm{.} \label{eq:PhaseRotLegendreDecomposition}
\end{equation}
In case this form of the rotation is inserted into the partial wave projection integrals of the general rotated waves $\tilde{A}_{\ell}$ (cf. equation (\ref{eq:ContAmbTrafoPW})), the following {\it mixing formula} emerges \cite{LegendreProductFormula}
\begin{equation}
 \boxed{\tilde{A}_{\ell} (W) = \sum_{k = 0}^{\infty} L_{k} (W) \hspace*{2pt} \sum_{m = \left| k - \ell \right|}^{ k + \ell } \left< k,0 ; \ell,0 | m , 0 \right>^{2} A_{m} (W) \mathrm{.} }  \label{eq:MixingFormulaBoxed}
\end{equation}
Here, $\left< j_{1}, m_{1} ; j_{2}, m_{2} | J , M \right>$ is just a usual Glebsch-Gordan coefficient. \newline
Some more remarks should be made on the formula (\ref{eq:MixingFormulaBoxed}): first of all, although it's derivation is not difficult, this author has (at least up to this point) not found this expression in the literature, at least in this particular form. However, mixing-phenomena have been pointed out for $\pi N$-scattering \cite{DeanLee} and for photoproduction \cite{Omelaenko}. \newline
Secondly, in can be seen quickly from the mixing formula that for angle-{\it in}dependent phases, i.e. when only the coefficient $L_{0}$ survives in the pa\-ra\-me\-tri\-za\-tion (\ref{eq:PhaseRotLegendreDecomposition}) of the rotation-functions, partial waves do not mix. Rather, in this case each partial wave is multiplied by $L_{0} (W) = e^{i \Phi (W)}$. However, once the phase $\Phi (W,\theta)$ carries even a weak angle-dependence, the expansion (\ref{eq:PhaseRotLegendreDecomposition}) directly becomes infinite and thus introduces contributions to an infinite partial wave set via the mixing-formula. There may be (a lot of) cases where the series (\ref{eq:PhaseRotLegendreDecomposition}) converges quickly and in these instances, it is safe to truncate the infinite equation-system (\ref{eq:MixingFormulaBoxed}) at some point. \newline
It has to be stated that the mixing under very general con\-ti\-nu\-um ambiguity transformations may lead to the mis-identification of resonance quantum numbers (reference \cite{Svarc2017} illustrates this fact on a toy-model example).

\section{Discrete ambiguities as con\-ti\-nu\-um ambiguity transformations} \label{sec:DiscreteAmbs}

In case of a polynomial-amplitude, i.e. a truncation of the infinite series (\ref{eq:BasicInfinitePWExpansion}) at some finite cutoff $L$, the amplitude decomposes into a product of linear factors \cite{Gersten}

\clearpage

\begin{equation}
A (W, \theta) = \sum_{\ell = 0}^{L} (2 \ell + 1) A_{\ell} (W) P_{\ell} (\cos \theta) \equiv \lambda \prod_{i = 1}^{L} \left( \cos \theta - \alpha_{i} \right) \mathrm{,} \label{eq:GerstenDecomposedScalarAmplitude}
\end{equation}
with a complex normalization proportional to the highest wave $\lambda \propto A_{L} (W)$. In case of a TPWA, one energy-dependent overall phase has to be fixed. This could be done, for instance, by choosing $\lambda$ real and positive: $\lambda = \left| \lambda \right|$. Sometimes it is also customary to fix the phase of the $S$-wave. \newline
Gersten \cite{Gersten} showed that discrete ambiguities in the TPWA can occur in case subsets of the roots $\left\{\alpha_{i}\right\}$ are complex conjugated. All combinatorial possibilities can be parametrized by a set of mappings $\bm{\uppi}_{\hspace*{0.035cm}p}$, the number of which rises exponentially with $L$:
\begin{equation}\bm{\uppi}_{\hspace*{0.035cm}p} \left(\alpha_{i}\right) := \begin{cases}
                    \alpha_{i} &\mathrm{,} \hspace*{3pt} \mu_{i} \left(p\right) = 0 \\
                    \alpha_{i}^{\ast} &\mathrm{,} \hspace*{3pt} \mu_{i} \left(p\right) = 1
                   \end{cases}\mathrm{,} \hspace*{5pt} p = \sum_{i = 1}^{L} \mu_{i} \left( p \right) 2^{(i-1)}\mathrm{,} \hspace*{5pt} p = 0,\ldots,(2^{L}-1) \mathrm{.} \label{eq:GerstenMapsScalarExample}
\end{equation}
In case these maps are applied, they yield a set of $2^{L}$ polynomial-amplitudes, which all have identical cross section:
\begin{equation}
A^{(p)} (W, \theta) = \lambda \prod_{i = 1}^{L} \left( \cos \theta - \bm{\uppi}_{\hspace*{0.035cm}p} \left[\alpha_{i}\right] \right) \equiv \sum_{\ell = 0}^{L} (2 \ell + 1) A^{(p)}_{\ell} (W) P_{\ell} (\cos \theta) \mathrm{.} \label{eq:GerstenTransformedScalarAmplitude}
\end{equation}
Since $\sigma_{0}$ is invariant under the discrete Gersten-ambiguities, these transformations can effectively only be rotations (because of $\left| A \right| = \sqrt{\sigma_{0}}$). More precisely, because one overall phase is fixed for all partial waves, discrete ambiguities can only be angle-dependent rotations. The corresponding rotation-functions are just fractions of two polynomial amplitudes
\begin{equation}
e^{i \varphi_{p} (W, \theta)} = \frac{A^{(p)} (W, \theta)}{A(W,\theta)} = \frac{\left( \cos \theta - \bm{\uppi}_{\hspace*{0.035cm}p} \left[\alpha_{1}\right] \right) \ldots \left( \cos \theta - \bm{\uppi}_{\hspace*{0.035cm}p} \left[\alpha_{L}\right] \right)}{\left( \cos \theta - \alpha_{1} \right) \ldots \left( \cos \theta - \alpha_{L} \right)} \mathrm{.} \label{eq:GerstenFormalismPhaseRotation}
\end{equation}
Therefore, discrete ambiguities mix partial waves, just as the general con\-ti\-nu\-um ambiguities do. Furthermore, the expression on the right-hand-side of (\ref{eq:GerstenFormalismPhaseRotation}) is explicitly an infinite series in $\cos \theta$. Thus, one may expect an infinite tower of rotated partial waves $\tilde{A}_{\ell}$ to be non-vanishing upon consideration of the mixing-formula (\ref{eq:MixingFormulaBoxed}). However, in this case of course the rotation fine-tunes exact cancellations in the results of the mixing for all higher partial waves $\tilde{A}_{\ell > L}$. \newline
Furthermore, Gersten \cite{Gersten} claims (without proof) that the root-conjugations { \it exhaust all possibilities} for discrete ambiguities of the TPWA. We have to state that we believe him.

\clearpage

The remainder of this proceeding is used to outline a numerical method that is {\it orthogonal} to the Gersten-formalism, but which can also substantiate this claim.

\section{Functional minimizations show exhaustiveness of Gersten-ambiguities} \label{sec:FunctMin}

We use the notation $x = \cos \theta$, introduce the complex rotation function $F (W,x) := e^{i \Phi(W,x)}$ and from now on drop the explicit energy $W$. The proposed numerical method assumes a truncated full amplitude $A(x)$ as a known input. Then, all possible functions $F(x)$ are scanned numerically for only those that satisfy the following two conditions:
\begin{itemize}
 \item[(I)] The complex solution-function $F (x)$ has to have modulus 1 for each value of $x$.
 \begin{equation}
 \left| F (x) \right|^{2} = 1 \mathrm{,} \hspace*{5pt} \forall x \in \left[ -1, 1 \right] \mathrm{.} \label{eq:PhaseModulus1}
 \end{equation}
 \item[(II)] The rotated amplitude $\tilde{A}(x)$, coming out of an amplitude $A(x)$ truncated at $L$, has to be truncated as well, i.e.
 \begin{equation}
 \tilde{A}_{L + k} = 0 \mathrm{,} \hspace*{5pt} \forall k = 1,\ldots,\infty \mathrm{.} \label{eq:RotAmplTruncatedAsWellRequireThis}
 \end{equation}
\end{itemize}
Formally, this {\it scanning}-procedure can be implemented by minimizing a suitably defined {\it functional} of $F(x)$:
\begin{align}
 &\bm{W} \left[ F(x) \right] := \sum_{x} \left(  \mathrm{Re}\left[F(x)\right]^{2}  + \mathrm{Im}\left[F(x)\right]^{2} - 1 \right)^{2} \nonumber \\ 
  & \hspace*{25.5pt} + \mathrm{Im} \left[  \frac{1}{2} \int_{-1}^{+1} dx \hspace*{1pt} F(x) A(x) \right]^{2}  
   + \sum_{k \geq 1} \Bigg\{ \mathrm{Re} \left[ \frac{1}{2} \int_{-1}^{+1} dx \hspace*{1pt} F(x) A(x) P_{L + k} (x) \right]^{2} \nonumber \\
  & \hspace*{25.5pt} + \mathrm{Im} \left[ \frac{1}{2} \int_{-1}^{+1} dx \hspace*{1pt} F(x) A(x) P_{L + k} (x) \right]^{2}  \Bigg\} \longrightarrow \mathrm{min} \mathrm{.} \label{eq:FunctProblem}
\end{align}
Here, the first term ensures the unimodularity of $F(x)$ (i.e. condition (I)), the second fixes a phase-convention on the $S$-wave $\tilde{A}_{0}$ and the big sum over $k$ sets all higher partial wave of the rotated amplitude to zero. \newline
It has to be clear that for practical numerical applications, the sums over $k$ and $x$ have to be finite, i.e. the former is cut off and the latter is defined on a grid of $x$-values. Also, a general function $F(x)$ is defined by an infinite amount of real degrees of freedom, which has to be made finite as well.

\clearpage

This can be achieved for instance by using a finite Legendre-expansion, i.e. a truncated version of equation (\ref{eq:PhaseRotLegendreDecomposition}) (with possibly large cutoff $\mathcal{L}_{\mathrm{cut}}$), or by discretizing $F(x)$ on a finite grid of points $\left\{x_{n}\right\} \in \left[-1, 1\right]$. More details on the numerical minimizations can be found in reference \cite{Wunderlich2017}. \newline
The only non-redundant solutions of this procedure are, in the end, the Gersten-rotation functions (\ref{eq:GerstenFormalismPhaseRotation}). Figures \ref{tab:FunctMinConvergencePlots1} and \ref{tab:FunctMinConvergencePlots2} illustrate this fact for the simple toy-model \cite{Wunderlich2017} (partial waves given in arbitrary units):
\begin{align}
 A(x) &= \sum_{\ell = 0}^{2} (2 \ell + 1) A_{\ell} P_{\ell} (x) = A_{0} + 3 A_{1} P_{1} (x) + 5 A_{2} P_{2} (x) \nonumber \\
  &= 5 + 3 (0.4 + 0.3 \hspace*{1pt} i) x + \frac{5}{2} (0.02 + 0.01 \hspace*{1pt} i) (3 x^{2} - 1) \mathrm{.} \label{eq:DefLmax2ToyModel}
\end{align}
This model is truncated at $L=2$. Thus it has two roots $\left(\alpha_{1},\alpha_{2}\right)$ and $2^{2} = 4$ Gersten-ambiguities. The latter are generated by four phase-rotation functions: $e^{i \varphi_{0} (x)} = 1$, $e^{i \varphi_{1} (x)}$, $e^{i \varphi_{2} (x)}$ and $e^{i \varphi_{3} (x)}$. Figures \ref{tab:FunctMinConvergencePlots1} and \ref{tab:FunctMinConvergencePlots2} demonstrate the convergence-process of the functional minimization towards a particular Gersten-rotation, for very general initial functions. The fact that always one of the four Gersten-rotations is found is {\it in}dependent of the choice of the initial function.

\section{Conclusions \& Outlook} \label{sec:Conclusions}

We have seen that general con\-ti\-nu\-um ambiguity transformations, as well as discrete Gersten-ambiguities, are in the end
manifestations of the same thing: angle-dependent phase-rotations. Therefore, they both mix partial waves. \newline
The rotations belonging to the Gersten-symmetries have the following defining property: they are the only rotations
which, if applied to an original truncated model, leave the truncation order $L$ untouched. In order to demonstrate this fact, 
a (possibly) new numerical method has been outlined capable of determining all con\-ti\-nu\-um ambiguity transformations satisfying 
pre-defined constraints. \newline
A possible further avenue of reserach may consist off the generalization of these findings to reactions with spin, for instance
pseudoscalar meson photoproduction. Here, the massive amount of new polarization data gathered over the last years have
renewed interest in questions of the uniqueness of partial wave decompositions. However, once one transitions to the case 
with spin, some open issues still exist, as have already been discussed during the workshop.

\clearpage

\begin{table*}[h]
\centering
\vspace*{-50pt}
\underline{\begin{Large}$e^{i \varphi_{0} (x)}$\end{Large}} \vspace*{5pt} \\
\includegraphics[width=0.2075\textwidth]{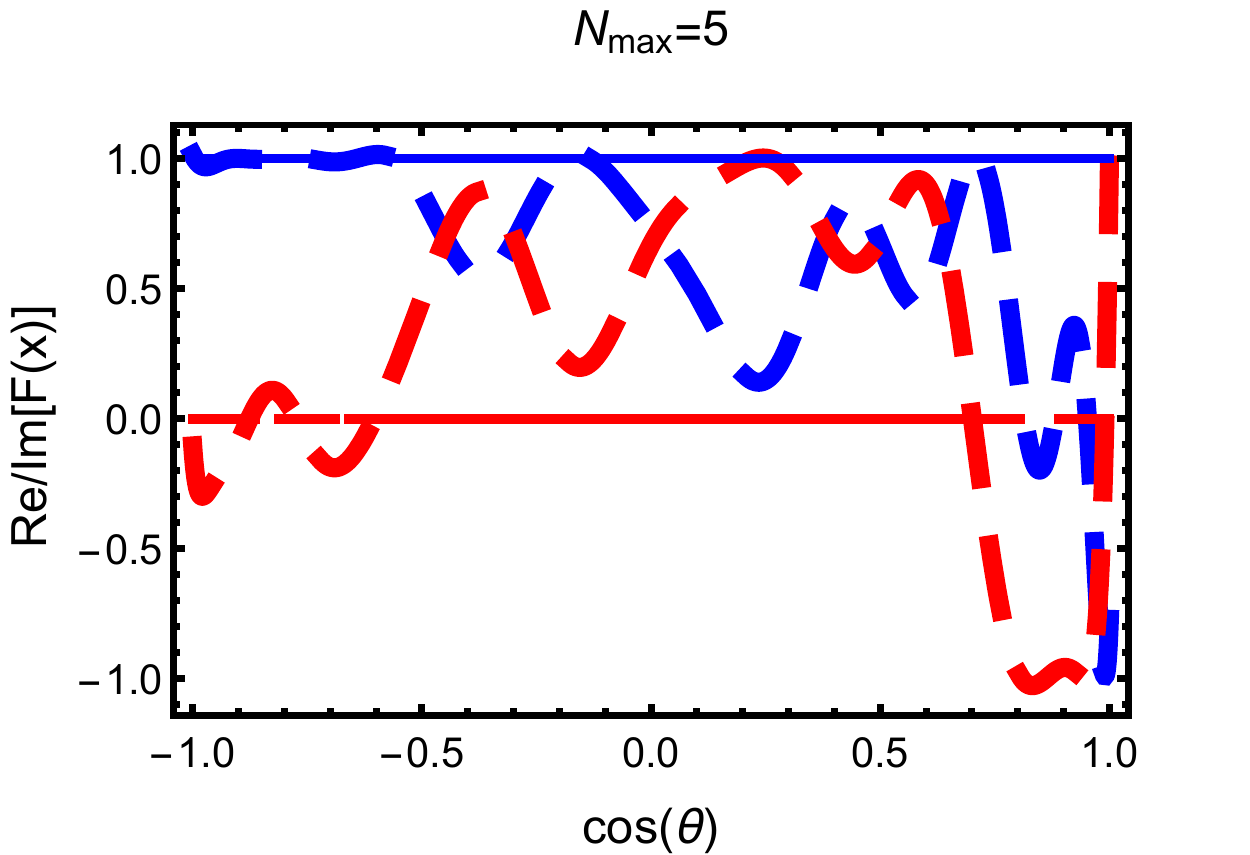}
\includegraphics[width=0.2075\textwidth]{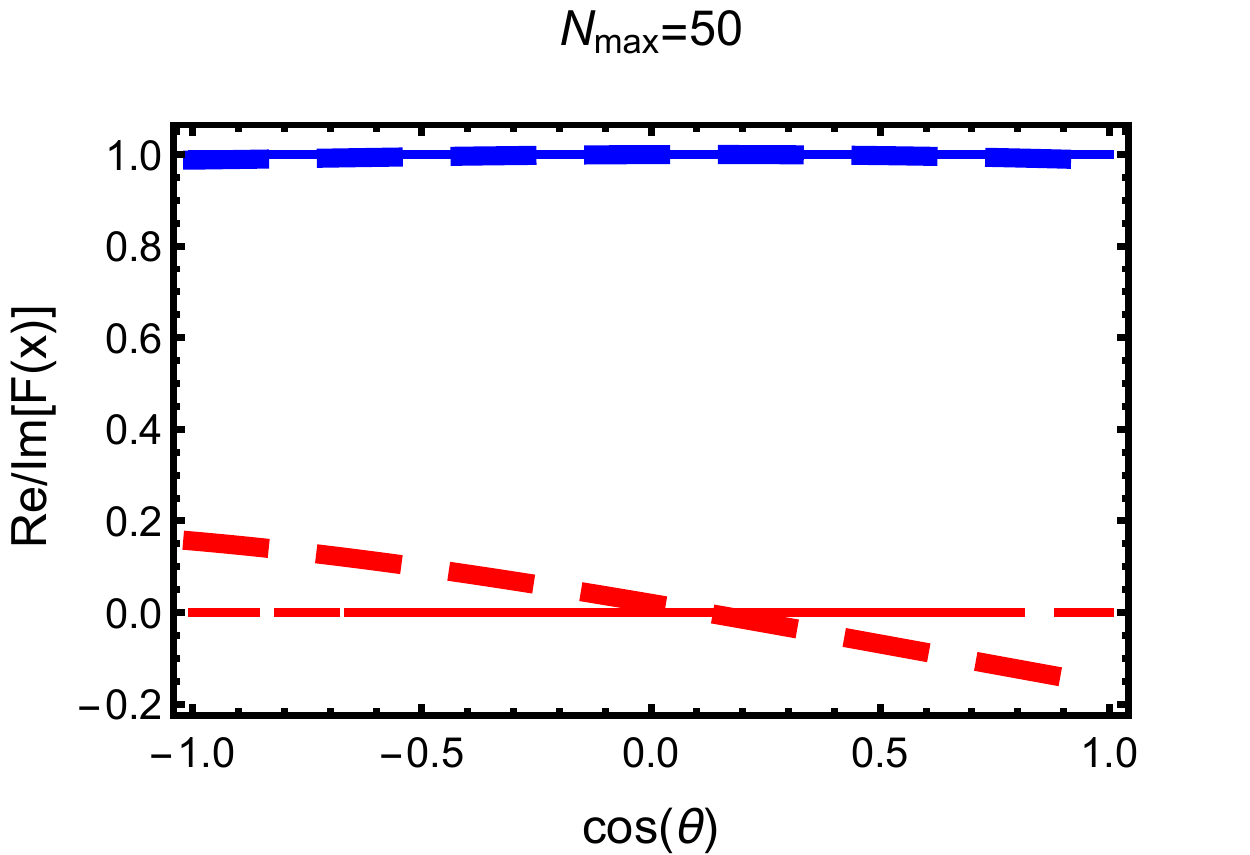}
\includegraphics[width=0.2075\textwidth]{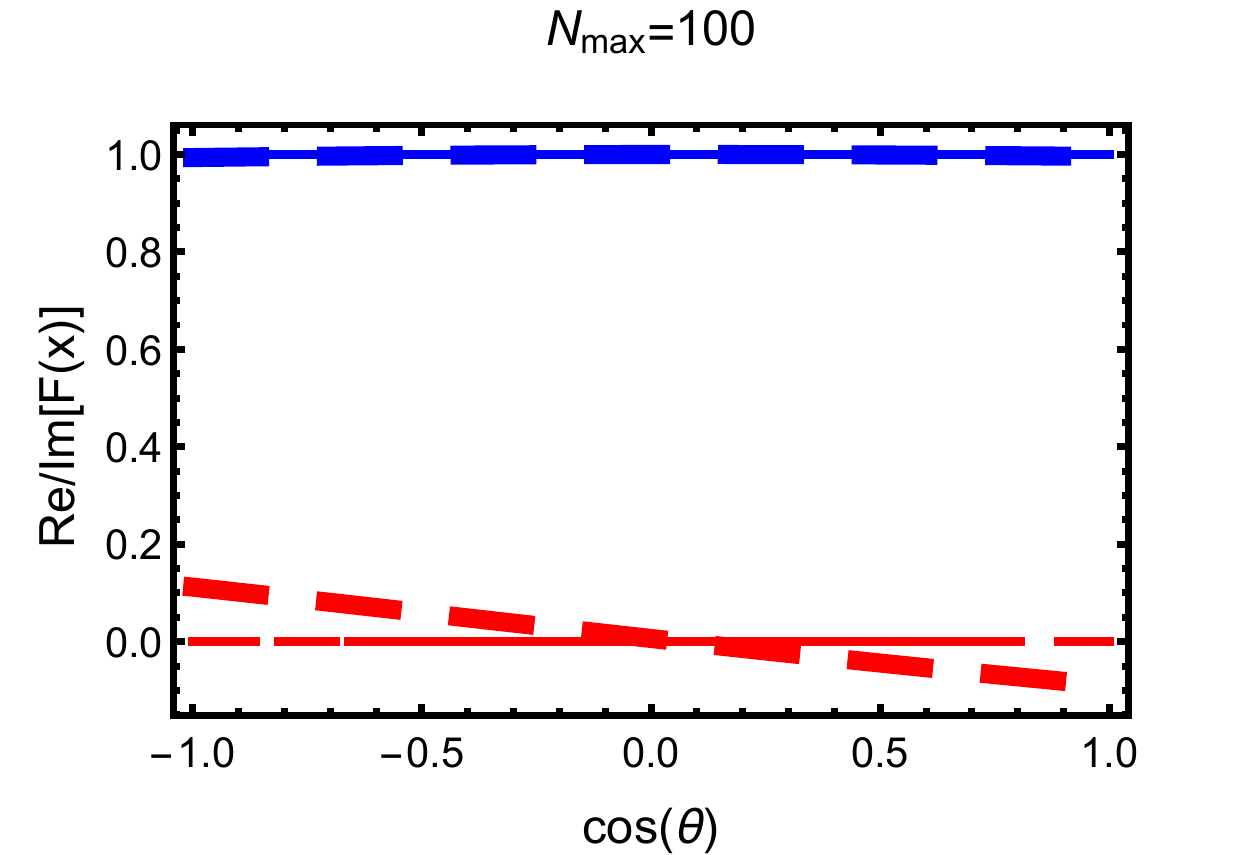}
\includegraphics[width=0.2075\textwidth]{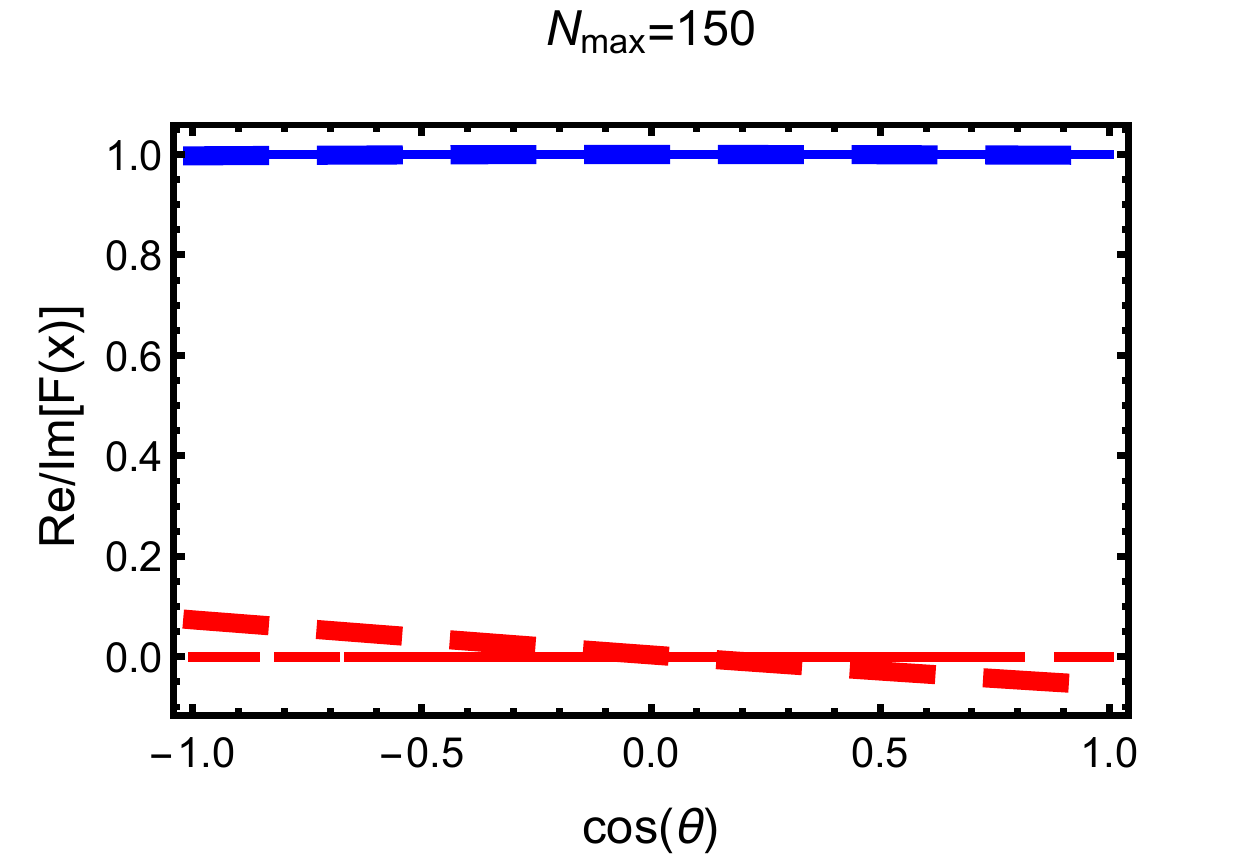} \\
\includegraphics[width=0.2075\textwidth]{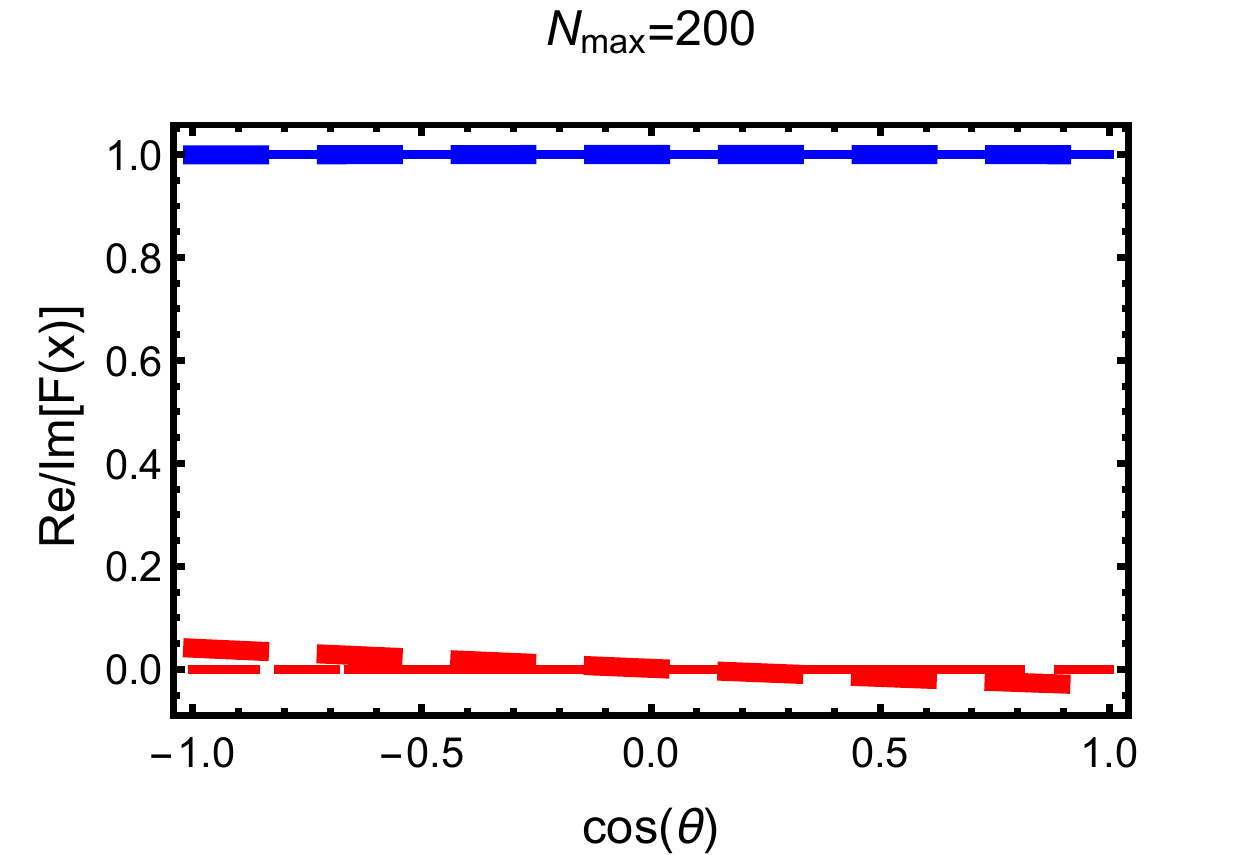}
\includegraphics[width=0.2075\textwidth]{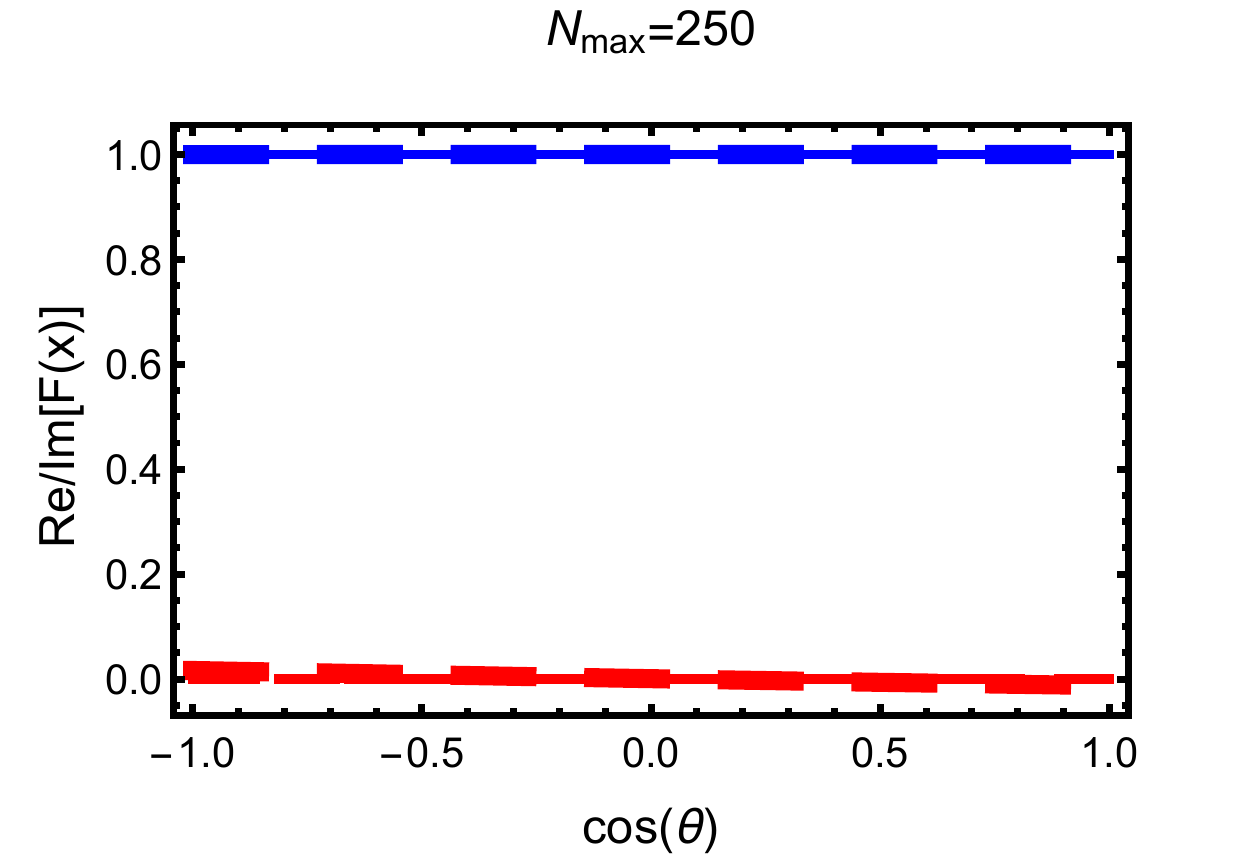}
\includegraphics[width=0.2075\textwidth]{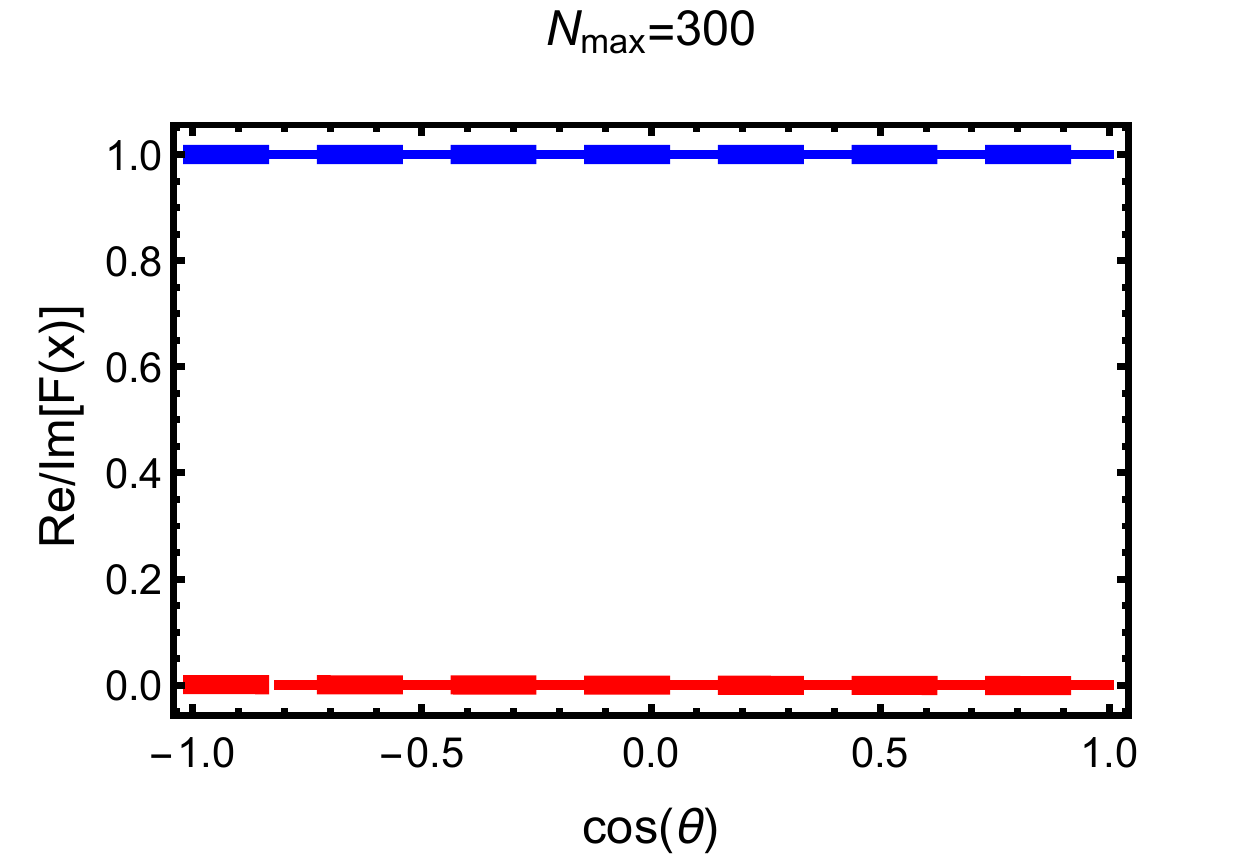}
\includegraphics[width=0.2075\textwidth]{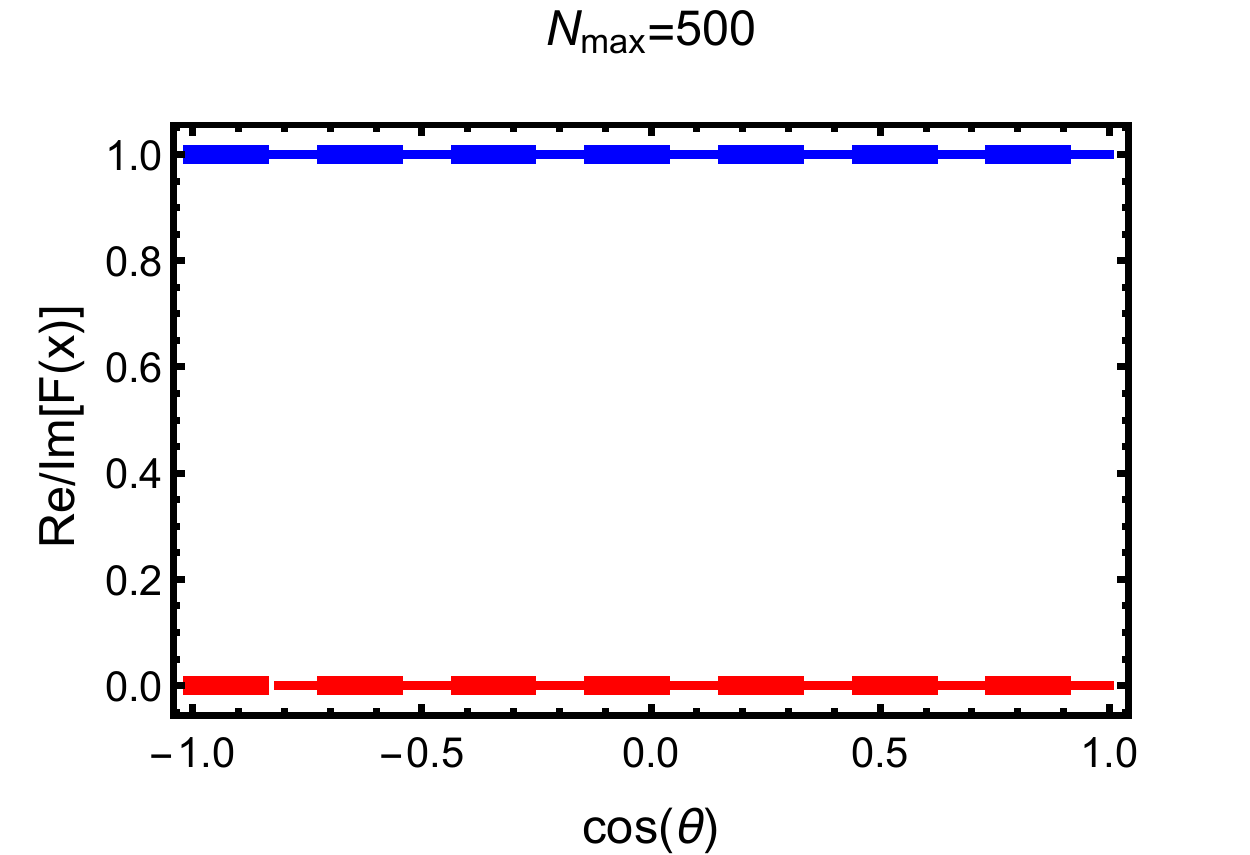} \\
\line(1,0){385} \\
\vspace*{5pt} \underline{\begin{Large}$e^{i \varphi_{1} (x)}$\end{Large}} \vspace*{5pt} \\
\includegraphics[width=0.2075\textwidth]{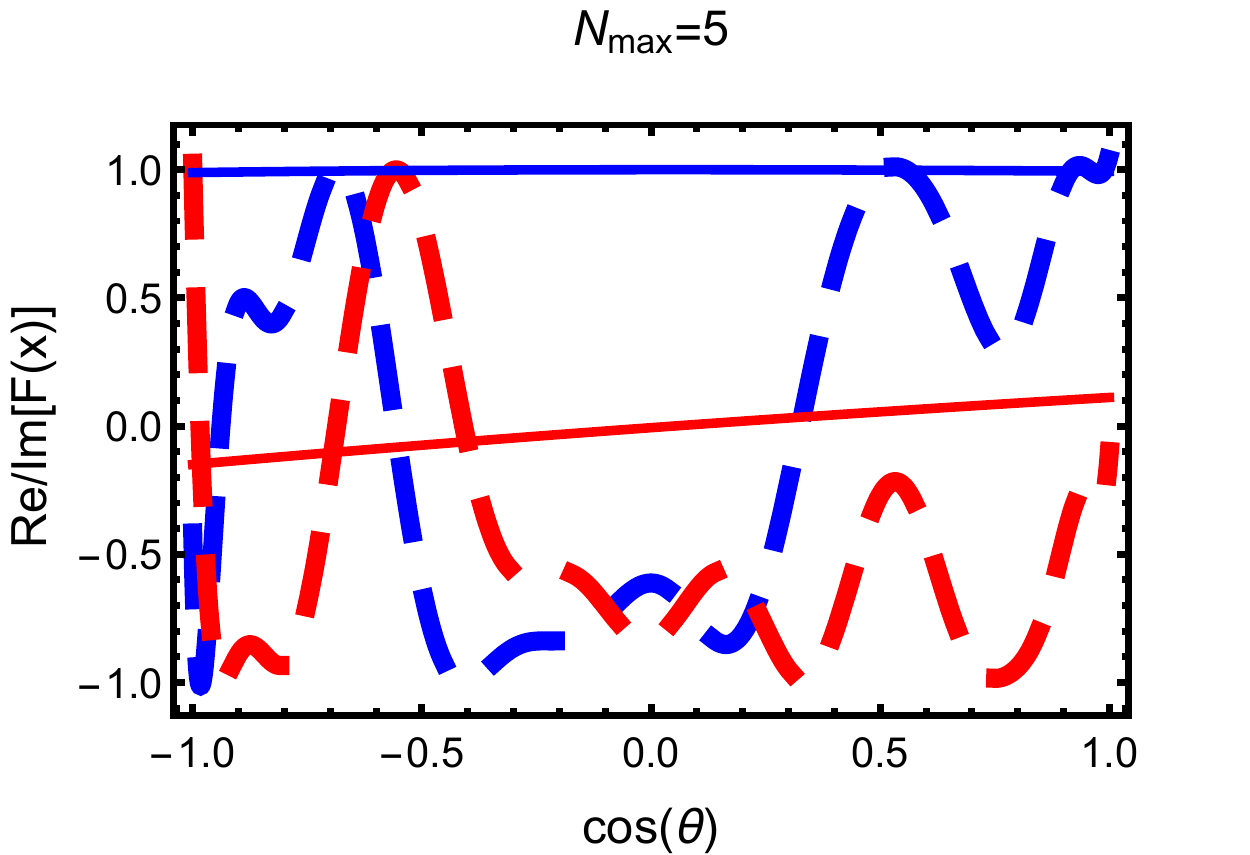}
\includegraphics[width=0.2075\textwidth]{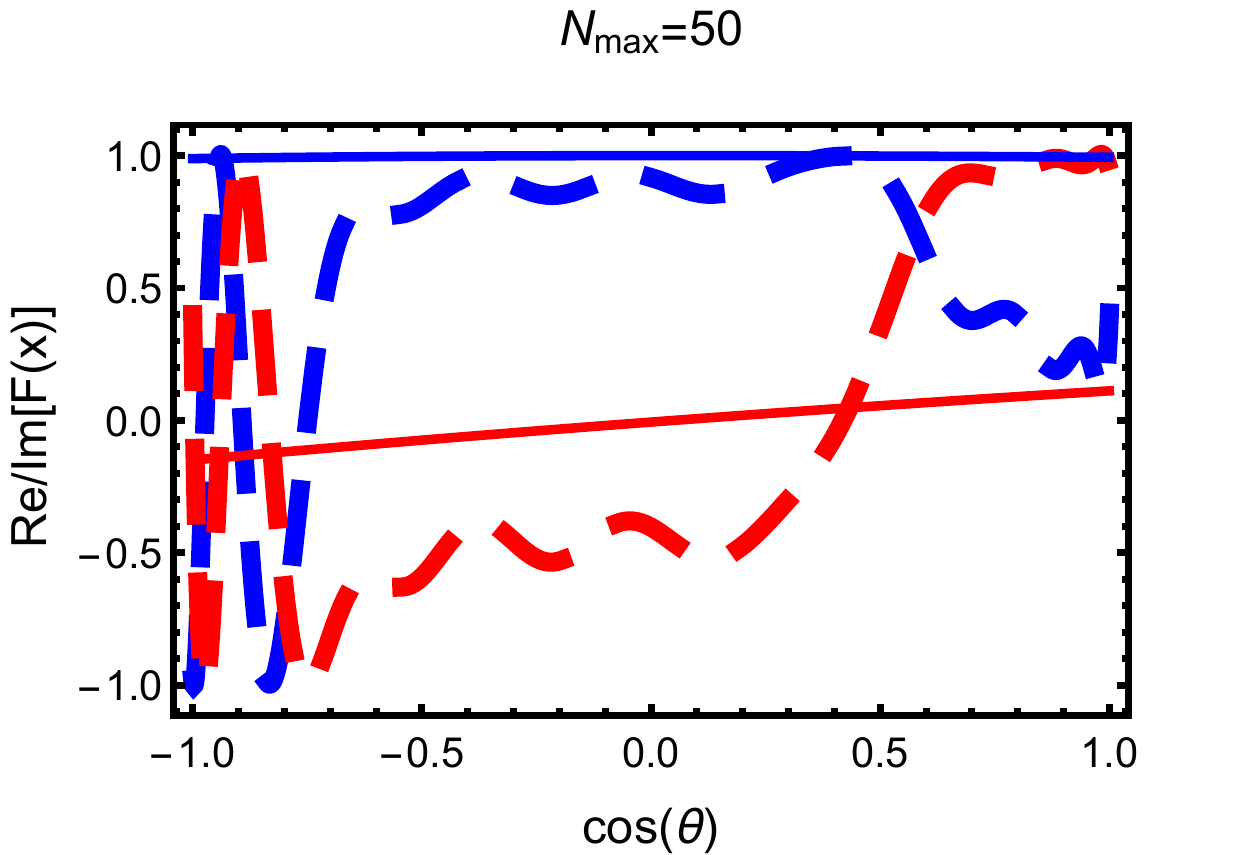}
\includegraphics[width=0.2075\textwidth]{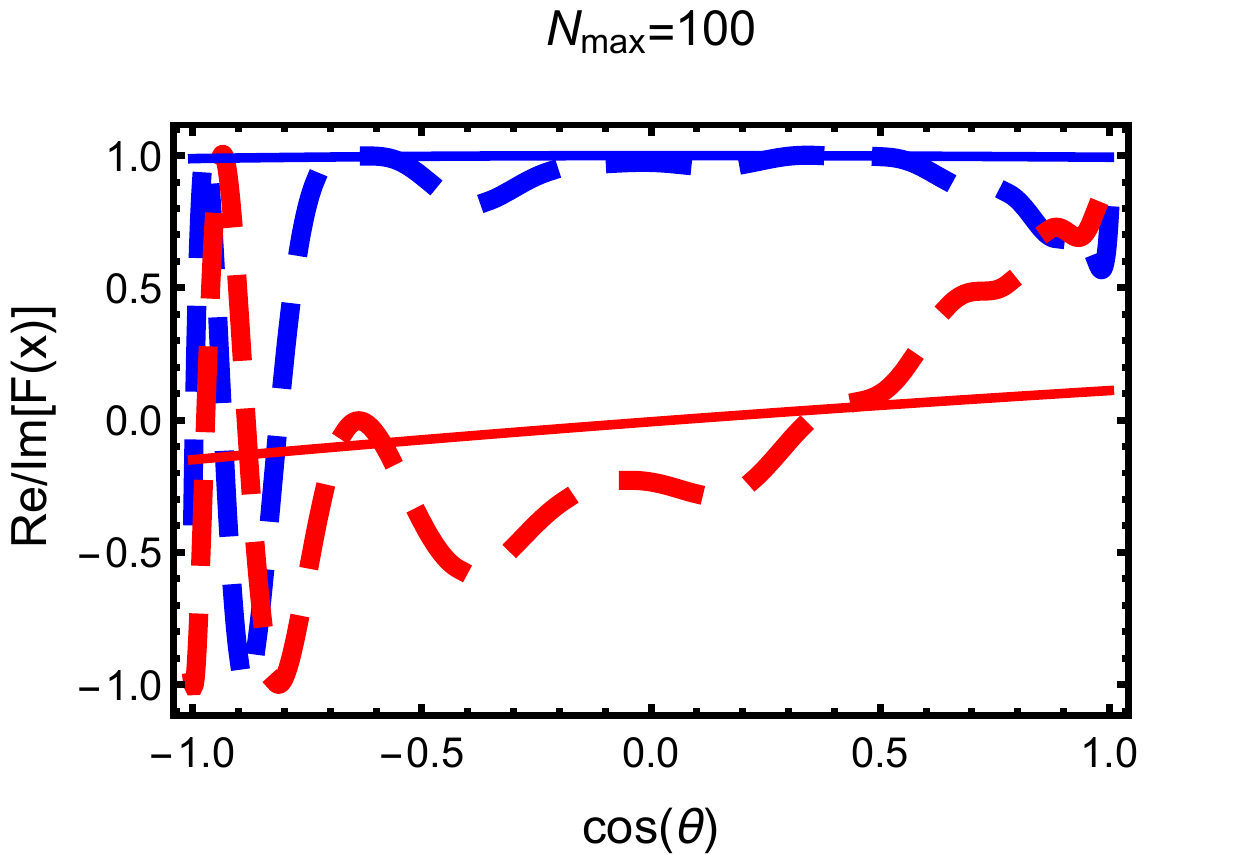}
\includegraphics[width=0.2075\textwidth]{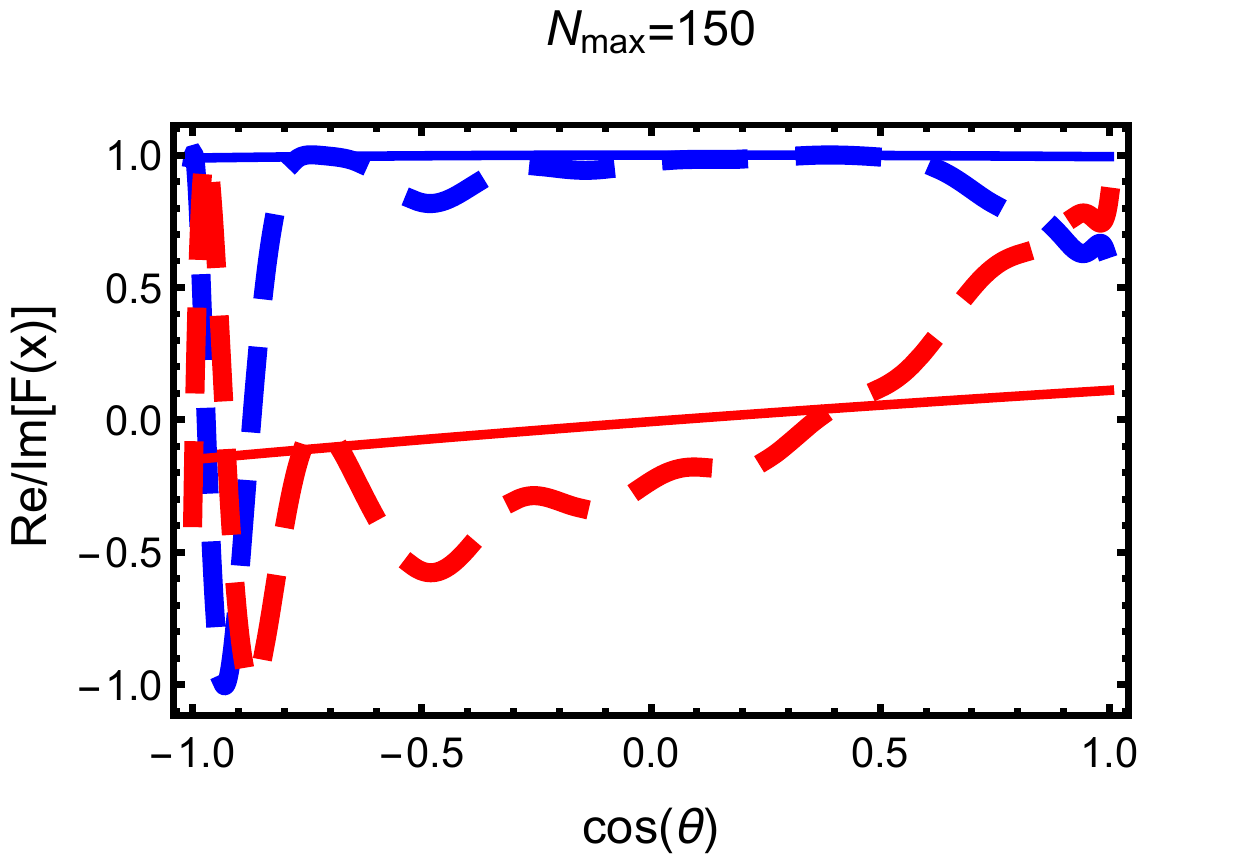} \\
\includegraphics[width=0.2075\textwidth]{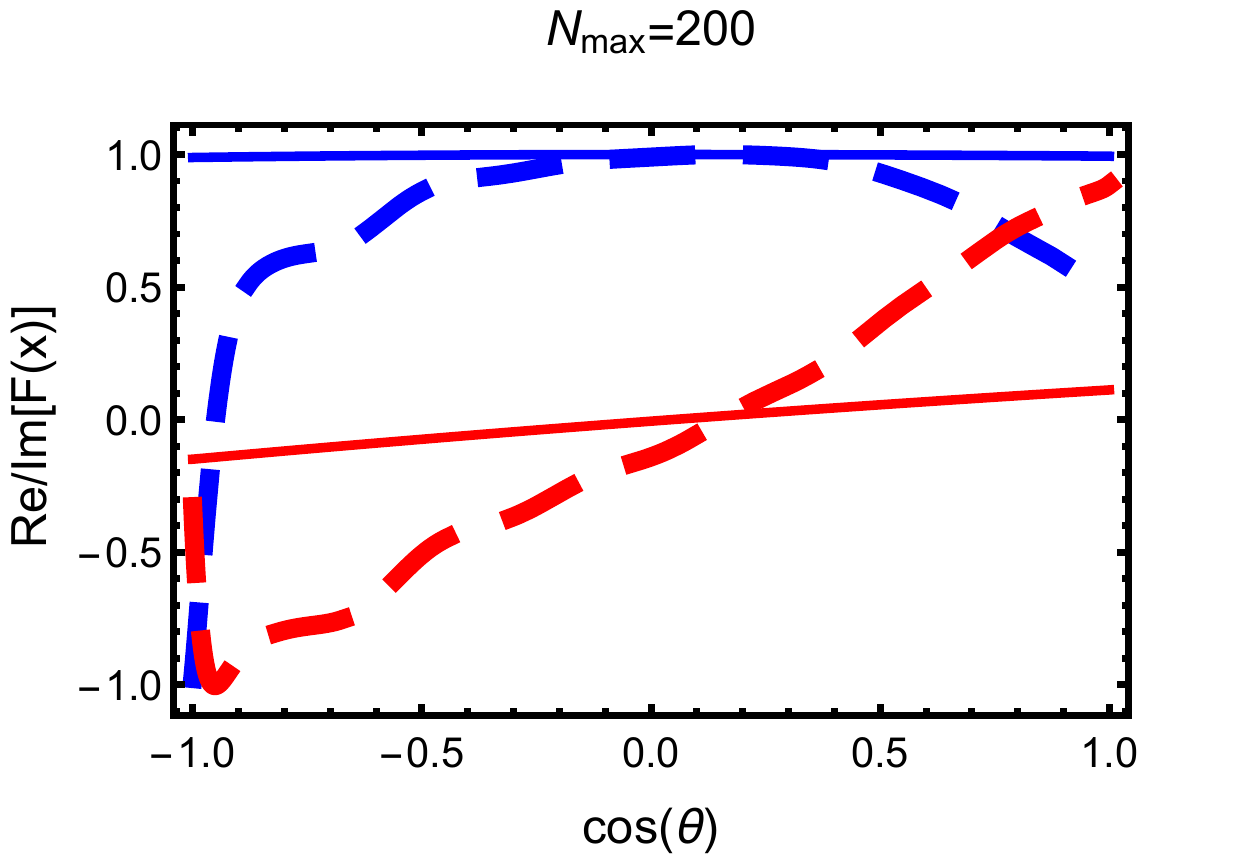}
\includegraphics[width=0.2075\textwidth]{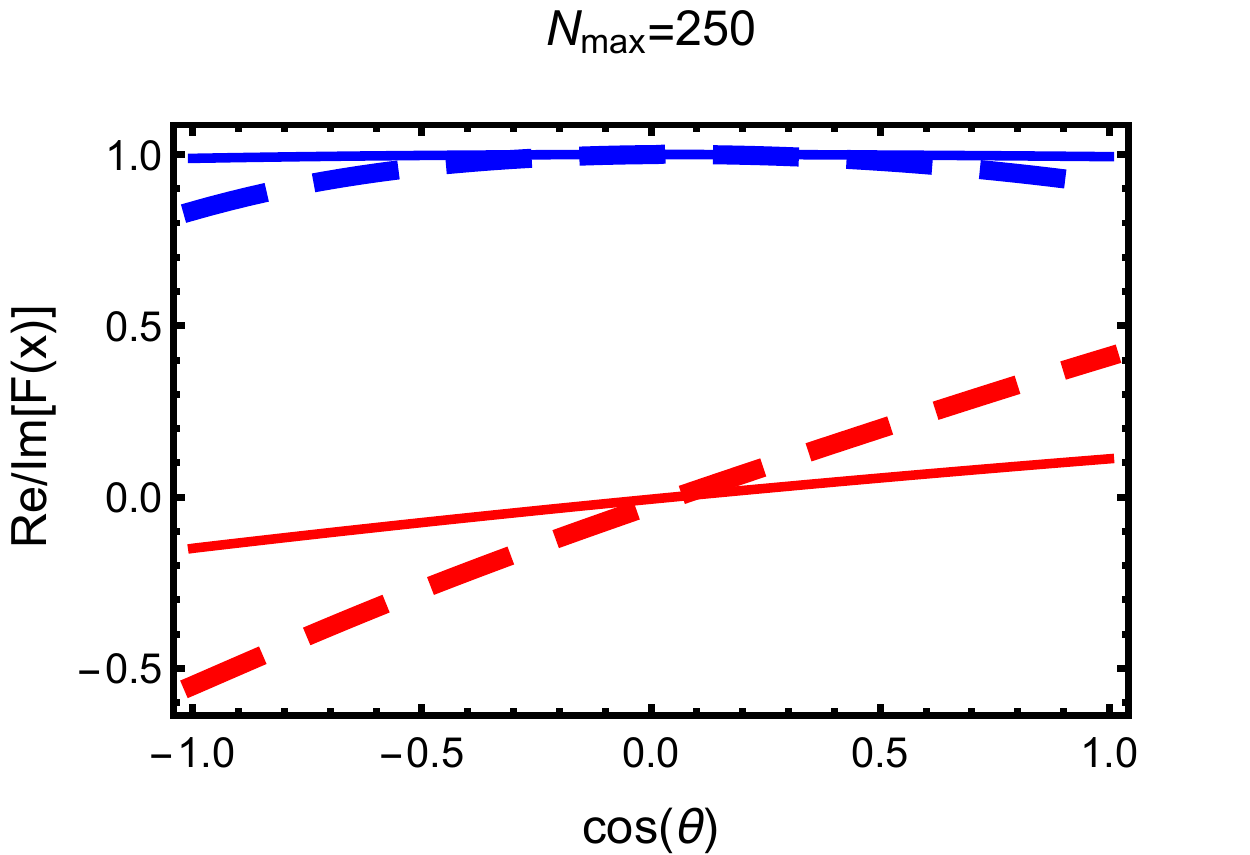}
\includegraphics[width=0.2075\textwidth]{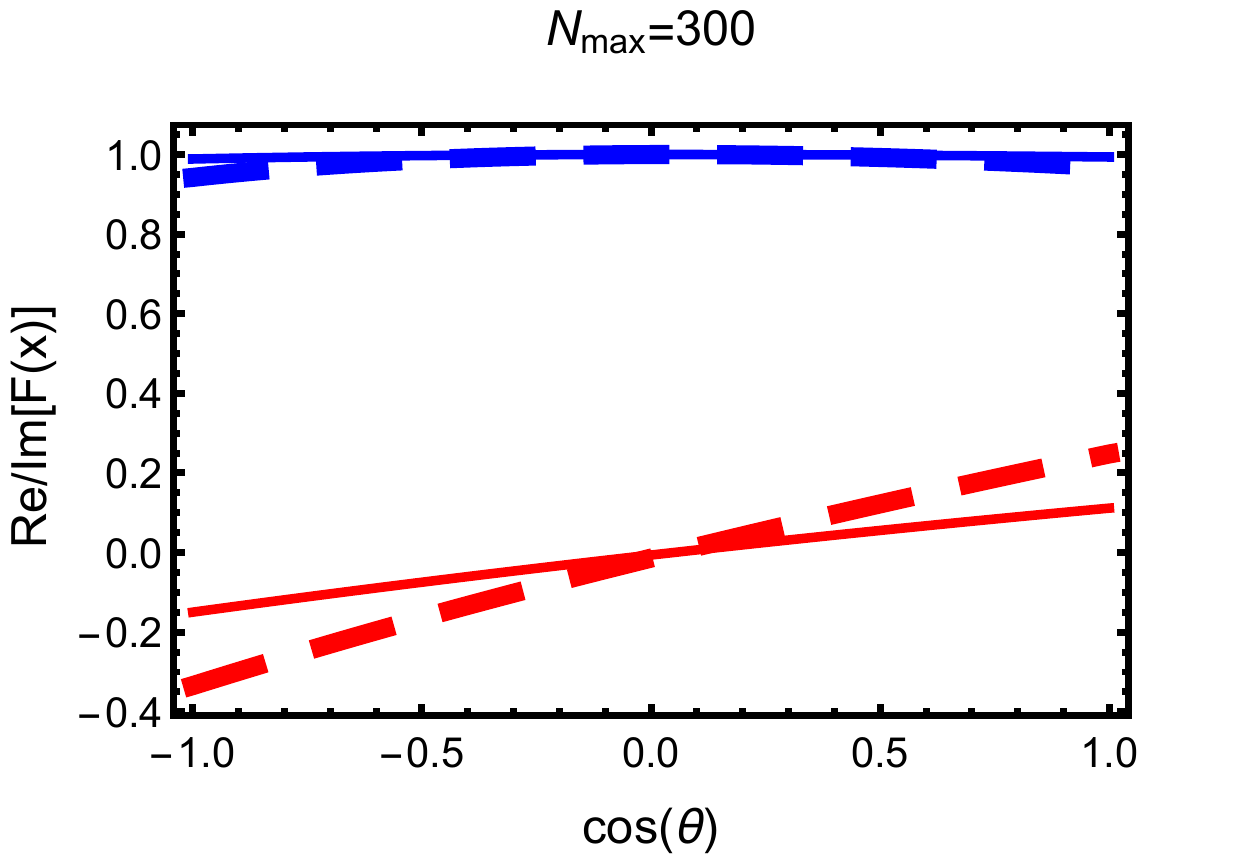}
\includegraphics[width=0.2075\textwidth]{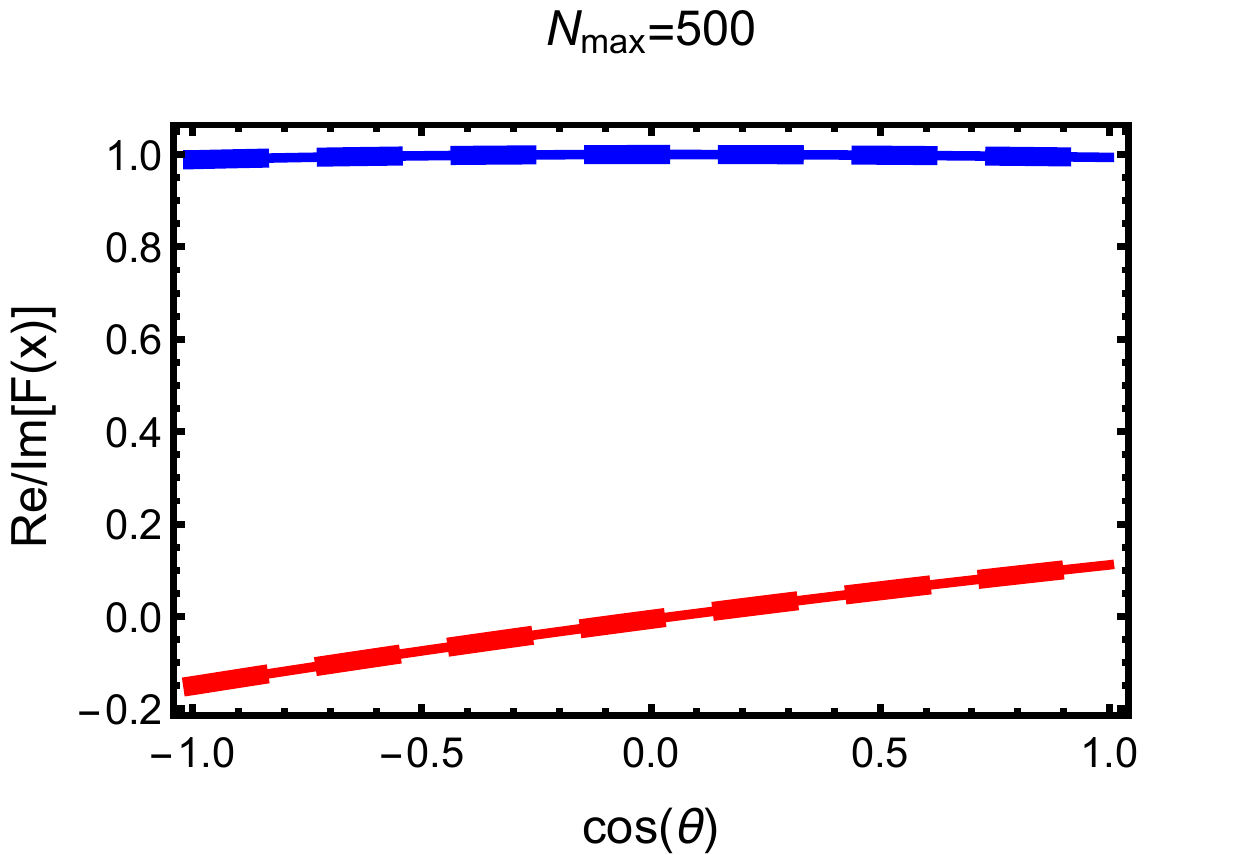} \\
% \caption{The convergence-process of the functional minimization procedure as described in the main text is demonstrated here. For the phase-rotations $e^{i \varphi_{0} (x)}$ and $e^{i \varphi_{1} (x)}$, generating the discrete ambiguities $\bm{\uppi}_{\hspace*{0.45pt}0}$ and $\bm{\uppi}_{\hspace*{0.45pt}1}$ of the toy-model (\ref{eq:DefLmax2ToyModel}), two randomly drawn initial functions have been picked from the applied ensemble. These initial conditions have, in the process of minimization, converged to these two respective phases. \newline
% Minimizations have been performed by starting always at the same initial function, but applying different numbers for the maximal number of iterations $N_{\mathrm{max}}$ of the minimizer, as indicated in the headers of the plots. Values range from $N_{\mathrm{max}} = 5$ (minimizer has barely changed the initial function) up to $N_{\mathrm{max}}=500$ (convergence-condition fulfilled for any of the minimizations). \newline
% In all plots, the real- and imaginary parts of the precise Gersten-ambiguity are drawn as blue and red solid lines. The results of the functional minimizations up to $N_{\mathrm{max}}$ are drawn as thick dashed lines, having the same color-coding for real- and imaginary parts. (color online)}
\captionof{figure}[fig]{The convergence of the functional minimization procedure is illustrated in these plots. For the discrete ambiguities $e^{i \varphi_{0} (x)}$ and $e^{i \varphi_{1} (x)}$ of the toy-model (\ref{eq:DefLmax2ToyModel}), two randomly drawn initial functions have been chosen from an applied ensemble of initial conditions in the search. These initial conditions then converged to these two respective Gersten-rotations. \newline
Results are shown for different values of the maximal number of iterations $N_{\mathrm{max}}$ of the minimizer, as indicated. Numbers range from $N_{\mathrm{max}} = 5$ up to $N_{\mathrm{max}}=500$. In all plots, the real- and imaginary parts of the precise Gersten-ambiguity are drawn as blue and red solid lines. The results of the functional minimizations up to $N_{\mathrm{max}}$ are drawn as thick dashed lines, having the same color-coding for real- and imaginary parts. (color online) \newline
These figures have already been published in reference \cite{Wunderlich2017}.}
\label{tab:FunctMinConvergencePlots1}
\end{table*}

\newpage

\begin{table*}[h]
\centering
\vspace*{-50pt}
\underline{\begin{Large}$e^{i \varphi_{2} (x)}$\end{Large}} \vspace*{5pt} \\
\includegraphics[width=0.2075\textwidth]{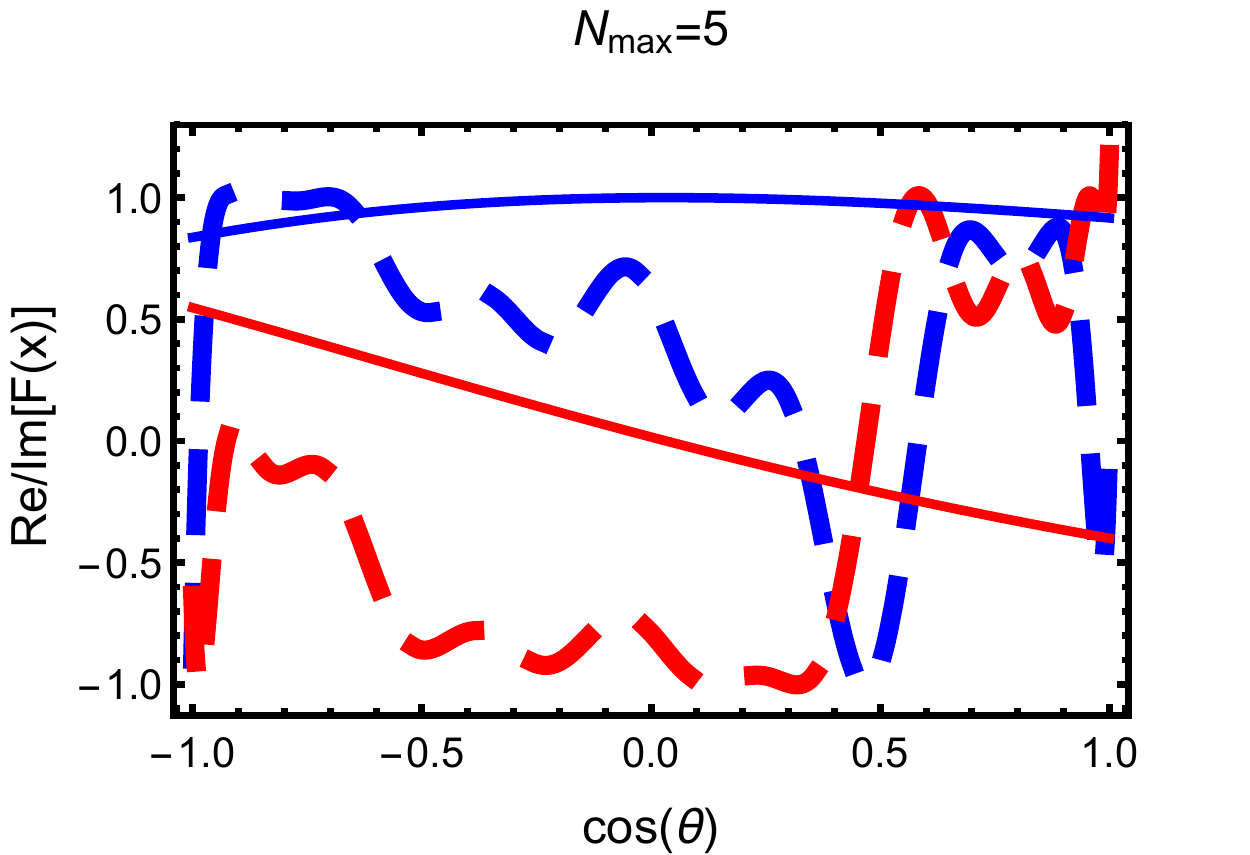}
\includegraphics[width=0.2075\textwidth]{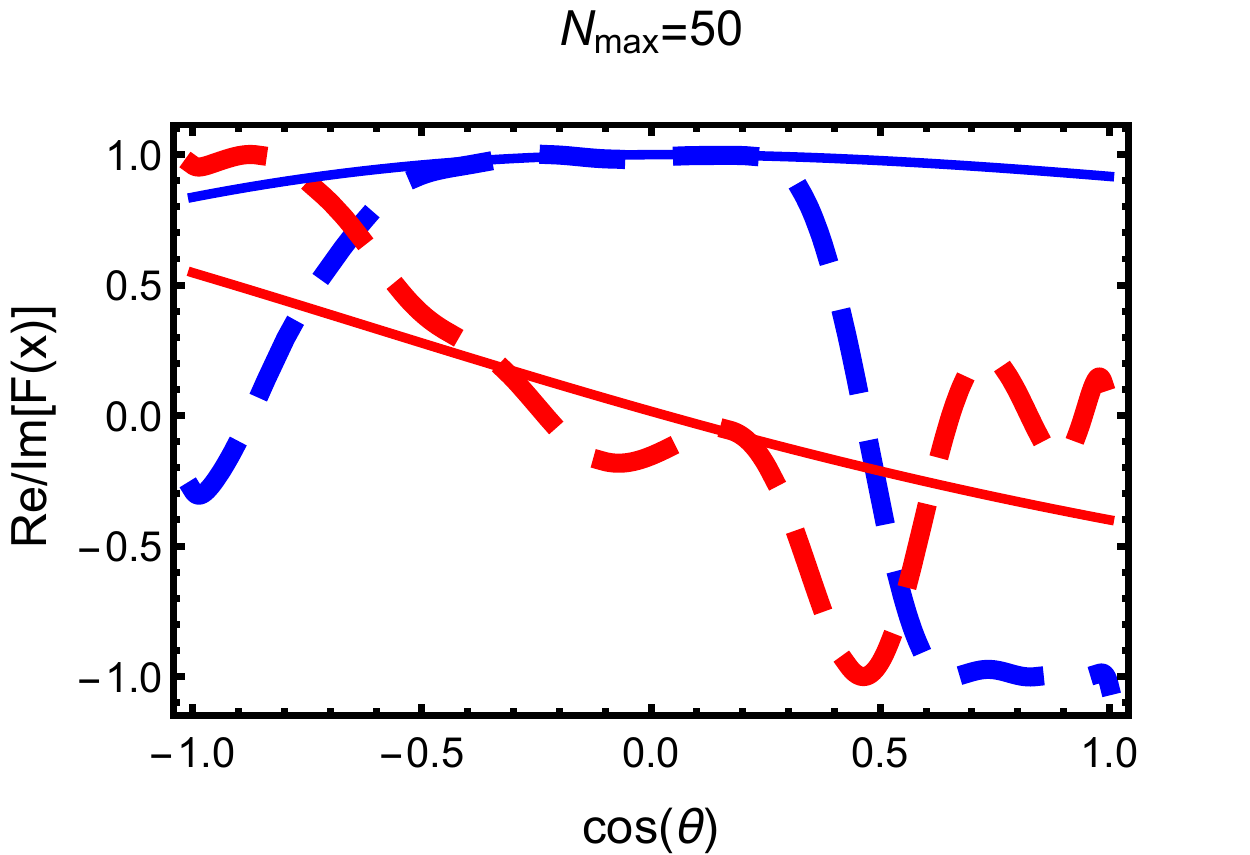}
\includegraphics[width=0.2075\textwidth]{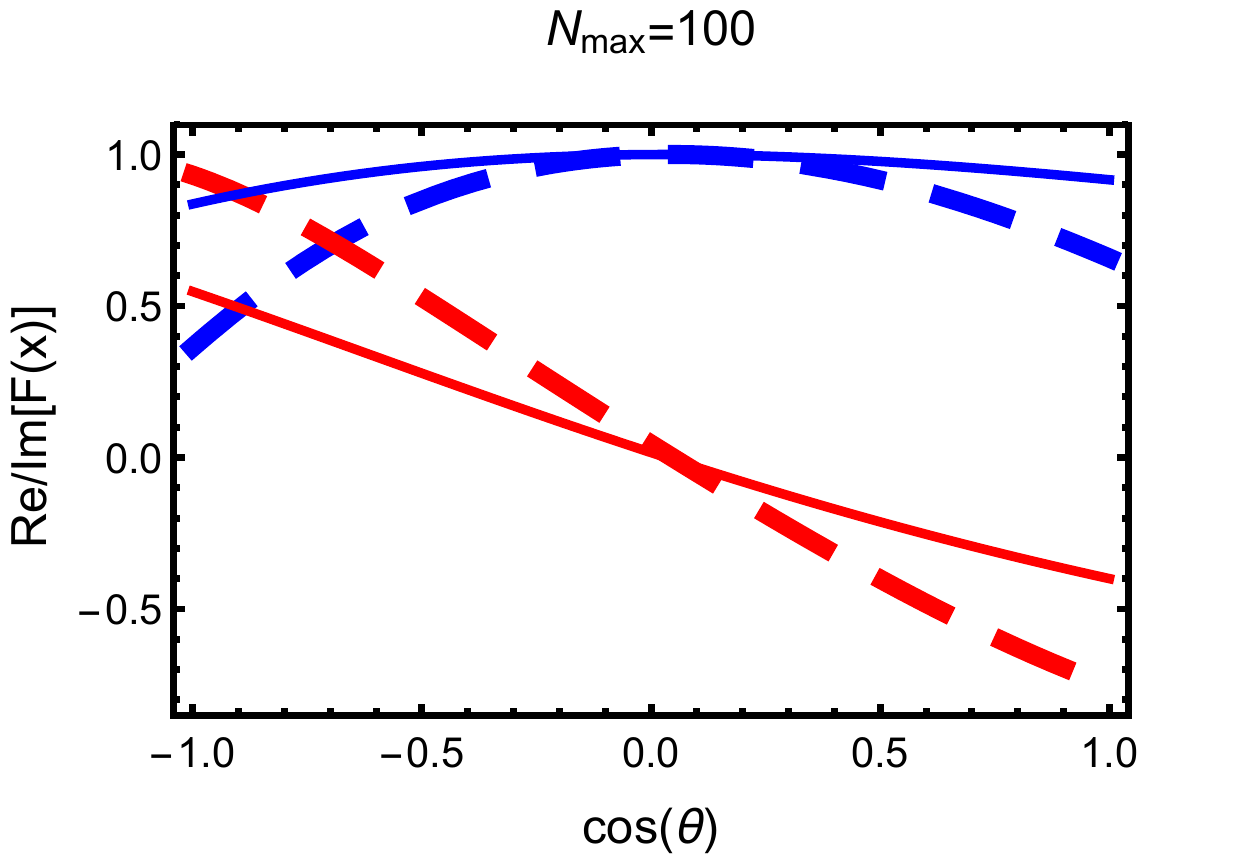}
\includegraphics[width=0.2075\textwidth]{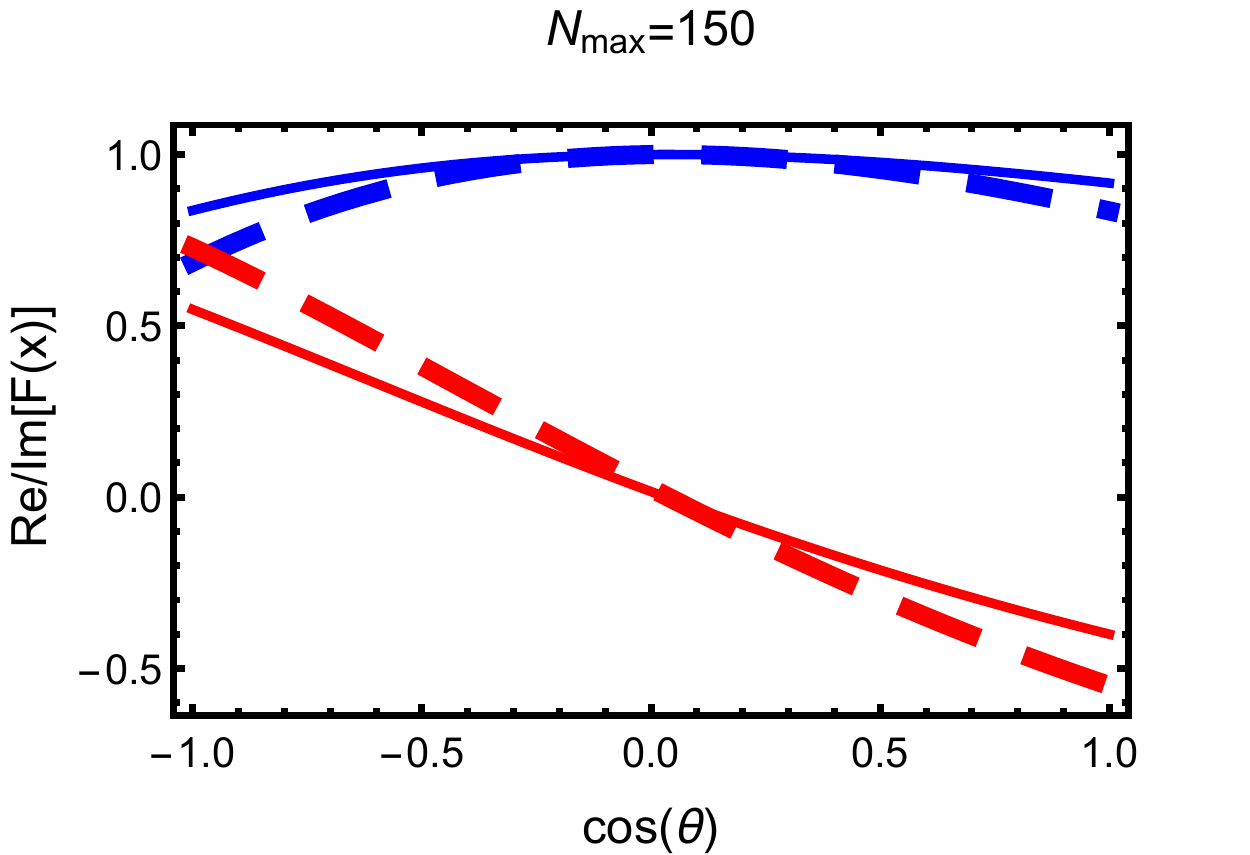} \\
\includegraphics[width=0.2075\textwidth]{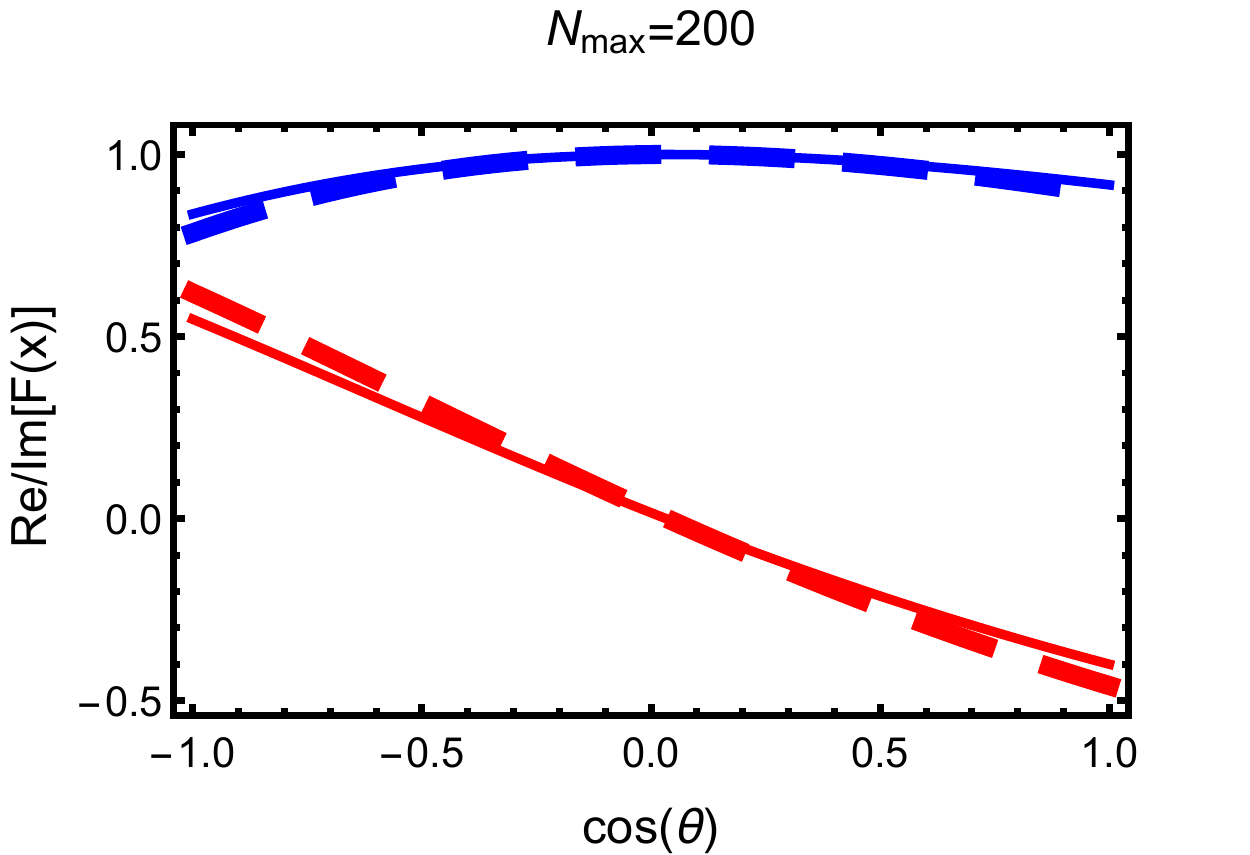}
\includegraphics[width=0.2075\textwidth]{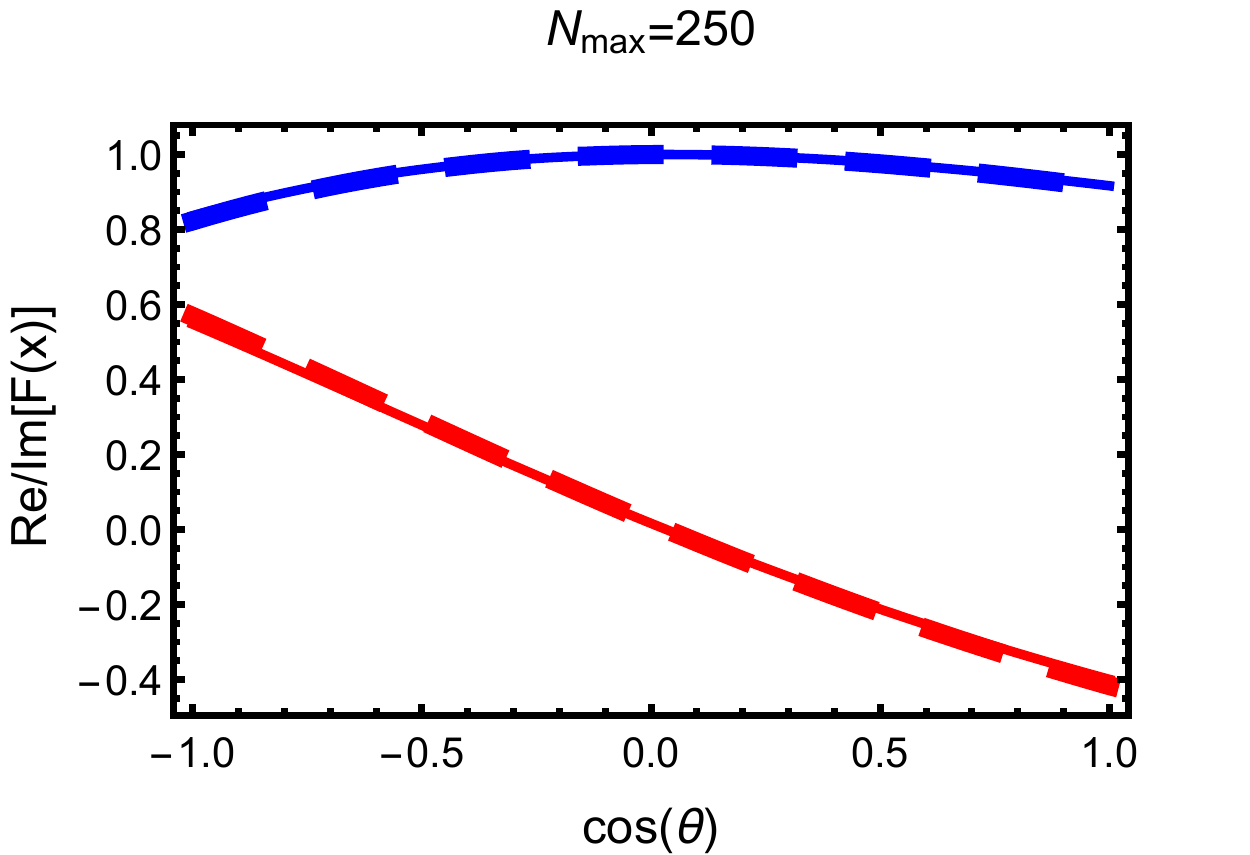}
\includegraphics[width=0.2075\textwidth]{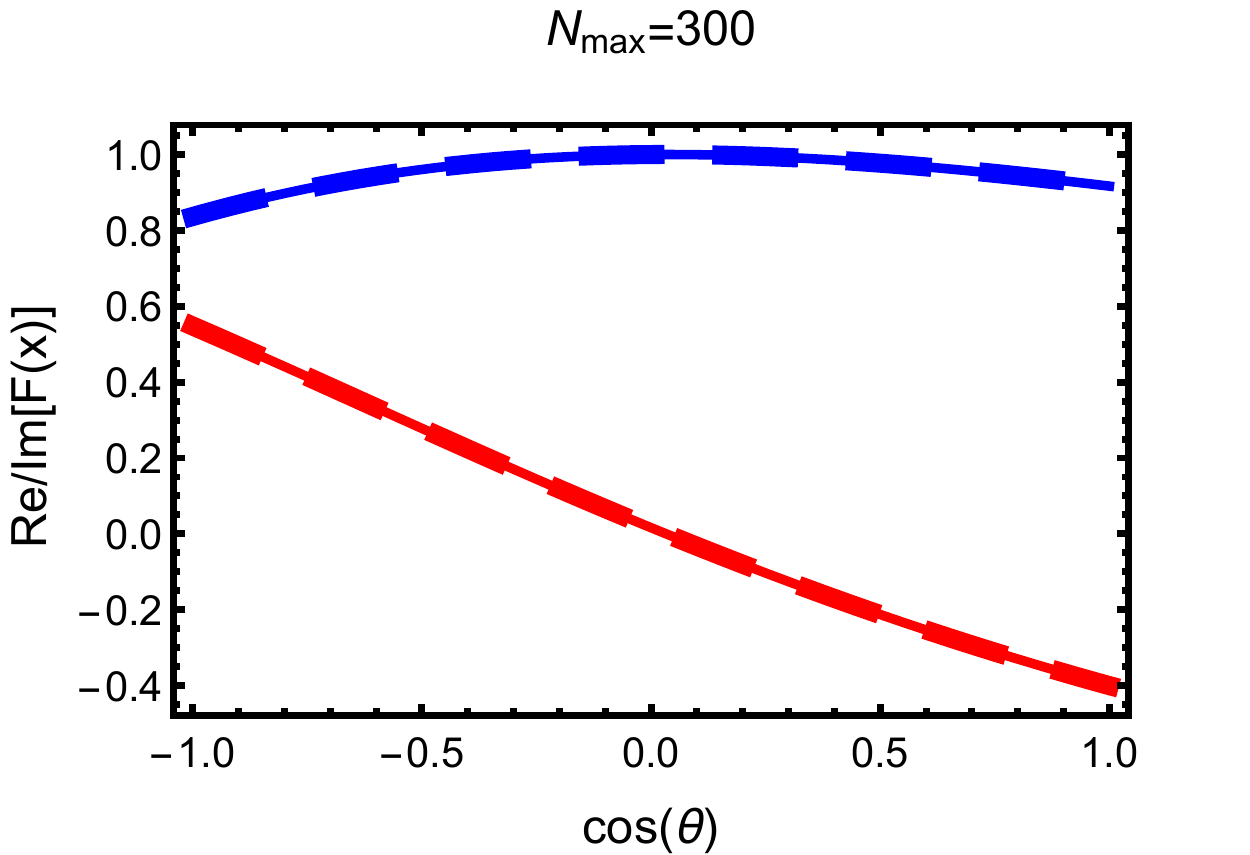}
\includegraphics[width=0.2075\textwidth]{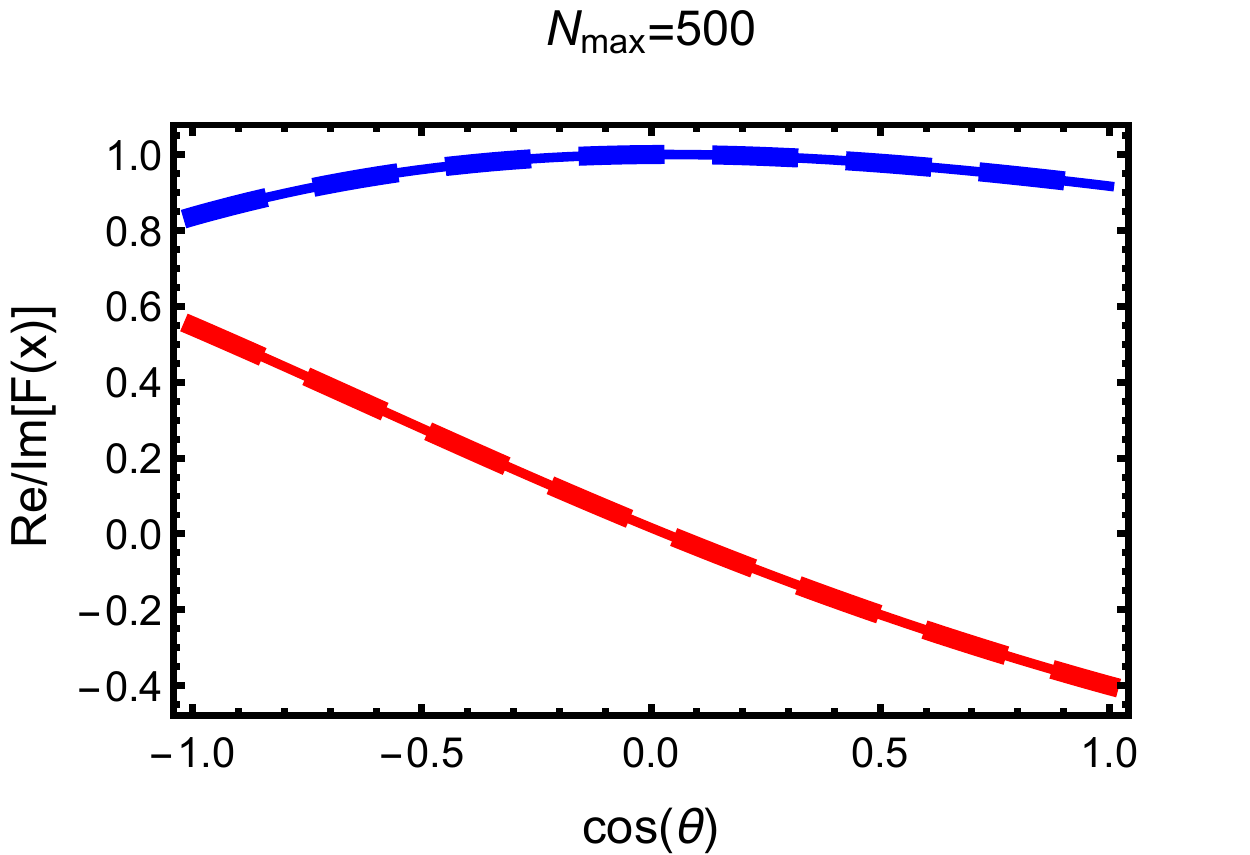} \\
\line(1,0){385} \\
\vspace*{5pt} \underline{\begin{Large}$e^{i \varphi_{3} (x)}$\end{Large}} \vspace*{5pt} \\
\includegraphics[width=0.2075\textwidth]{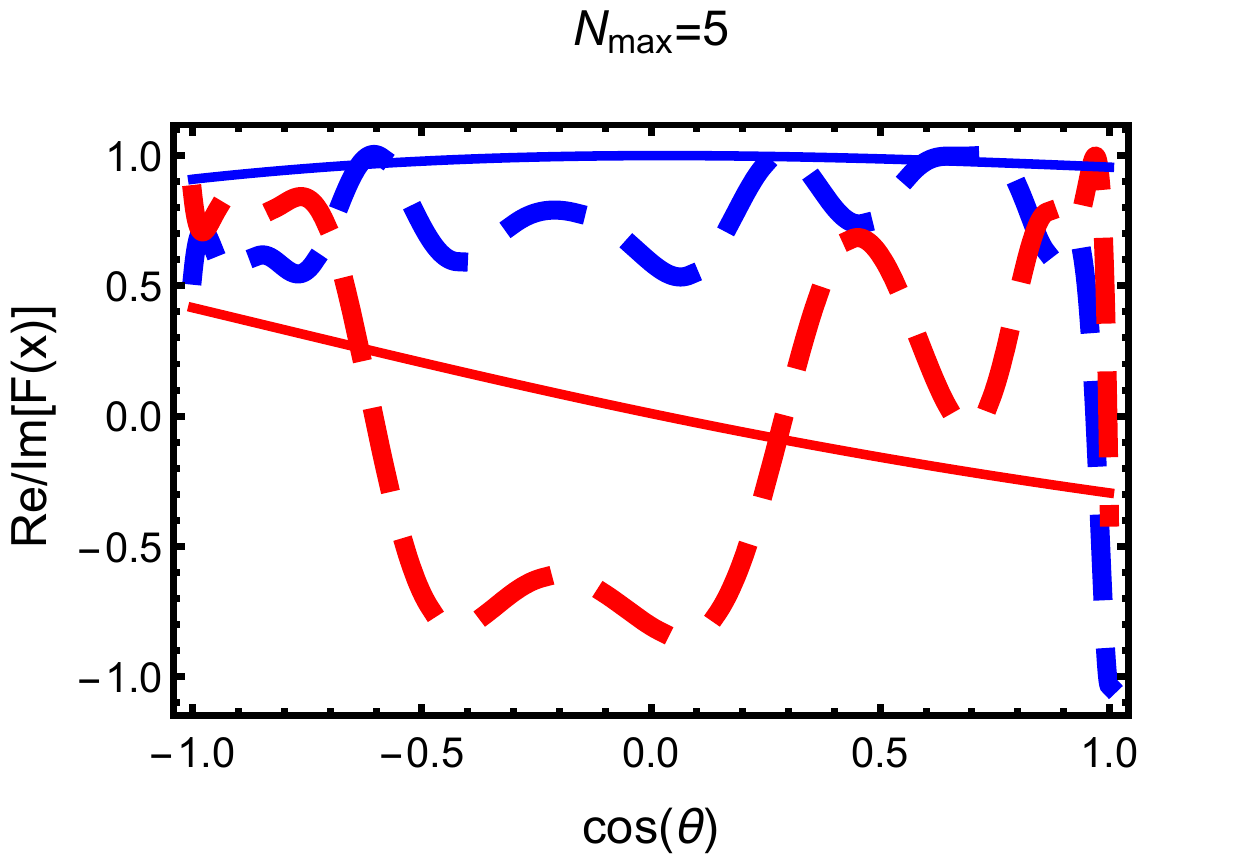}
\includegraphics[width=0.2075\textwidth]{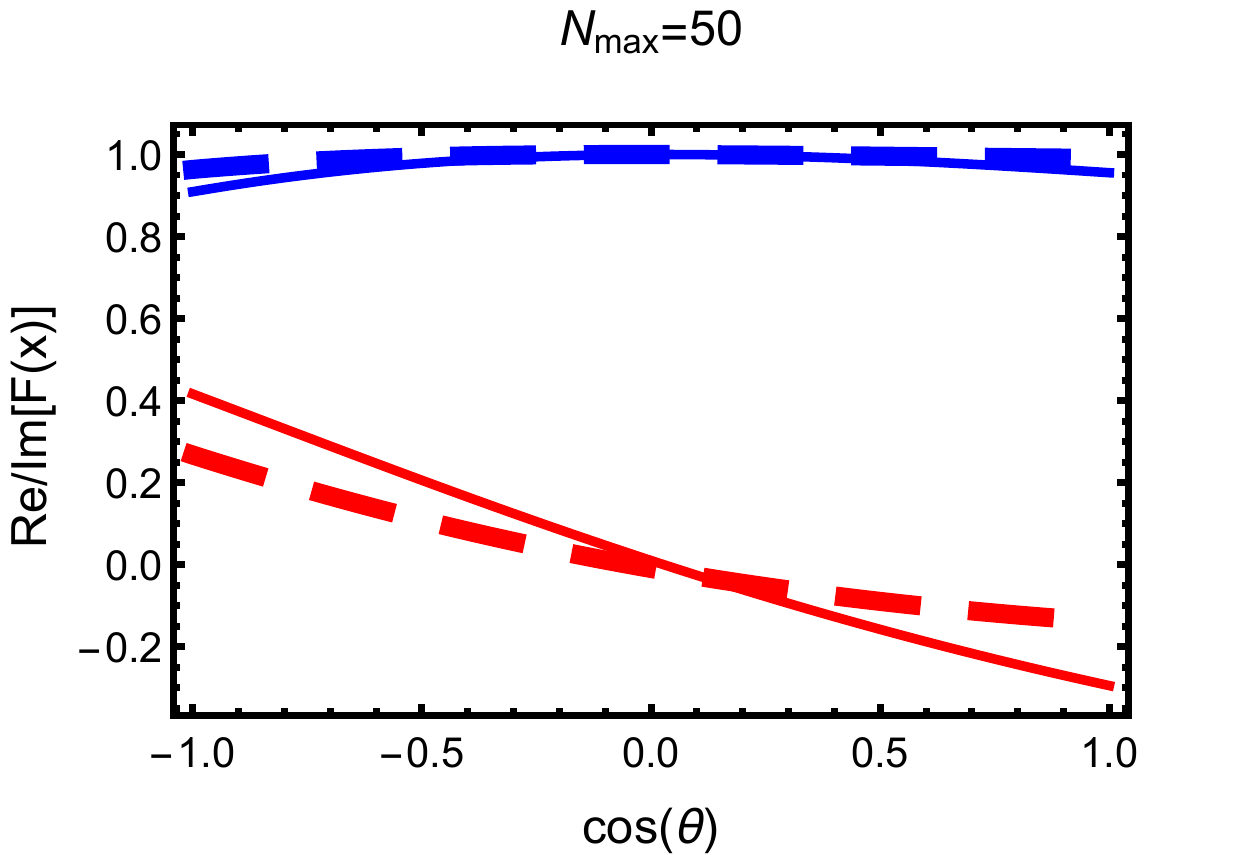}
\includegraphics[width=0.2075\textwidth]{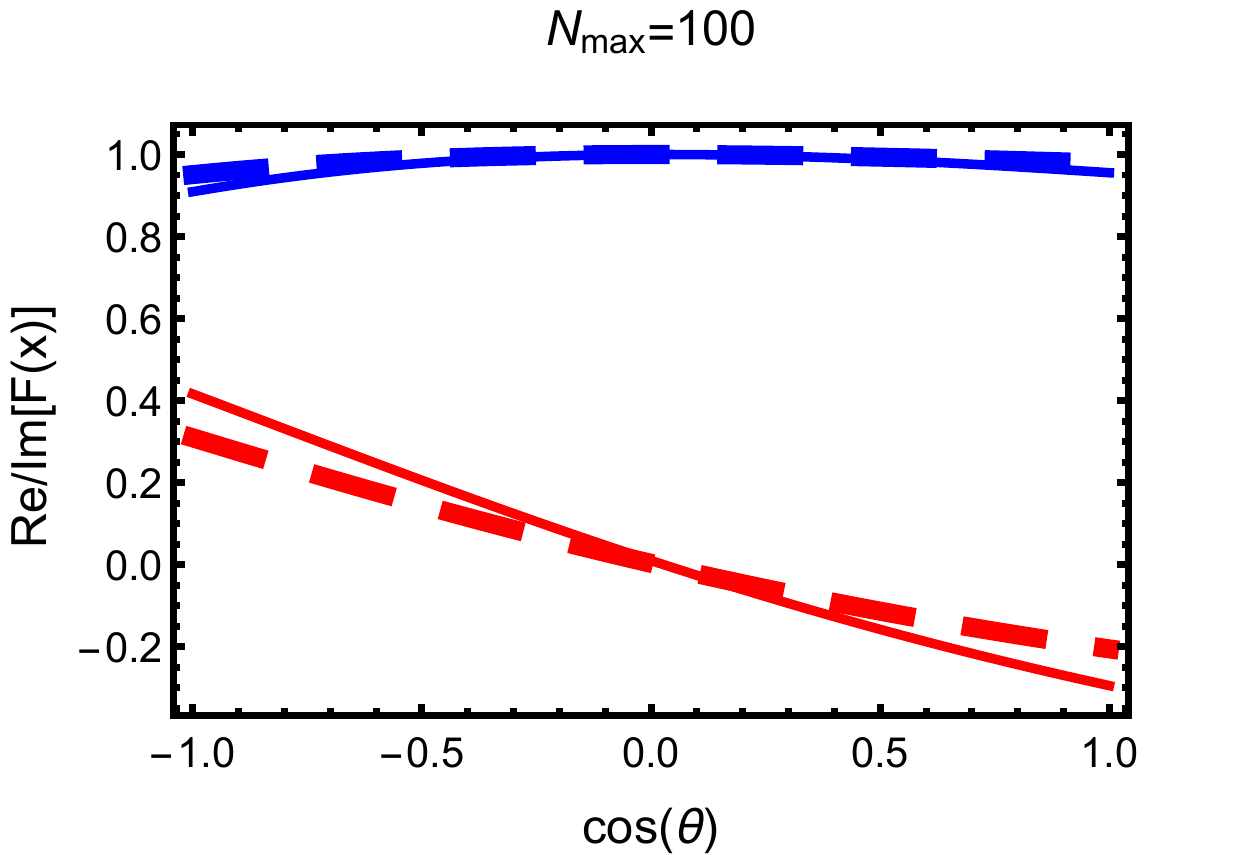}
\includegraphics[width=0.2075\textwidth]{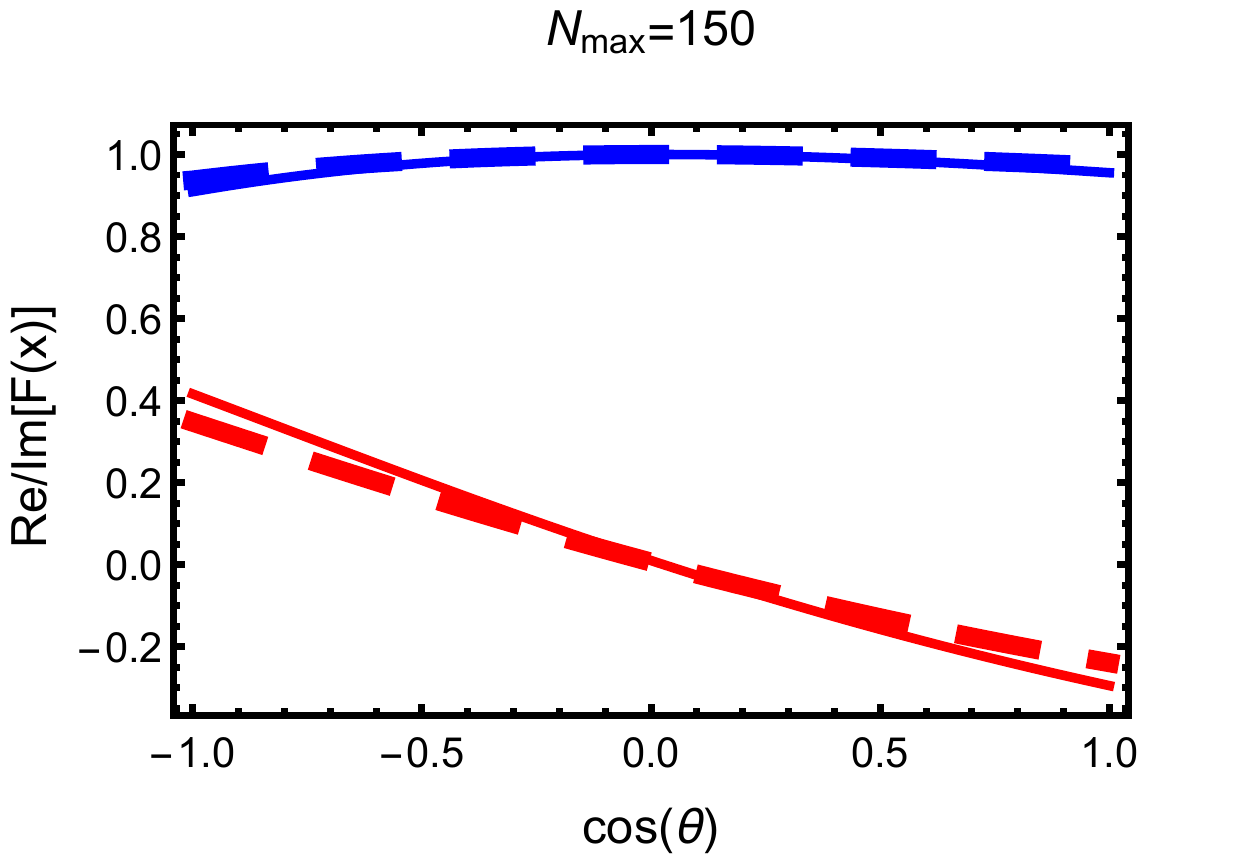} \\
\includegraphics[width=0.2075\textwidth]{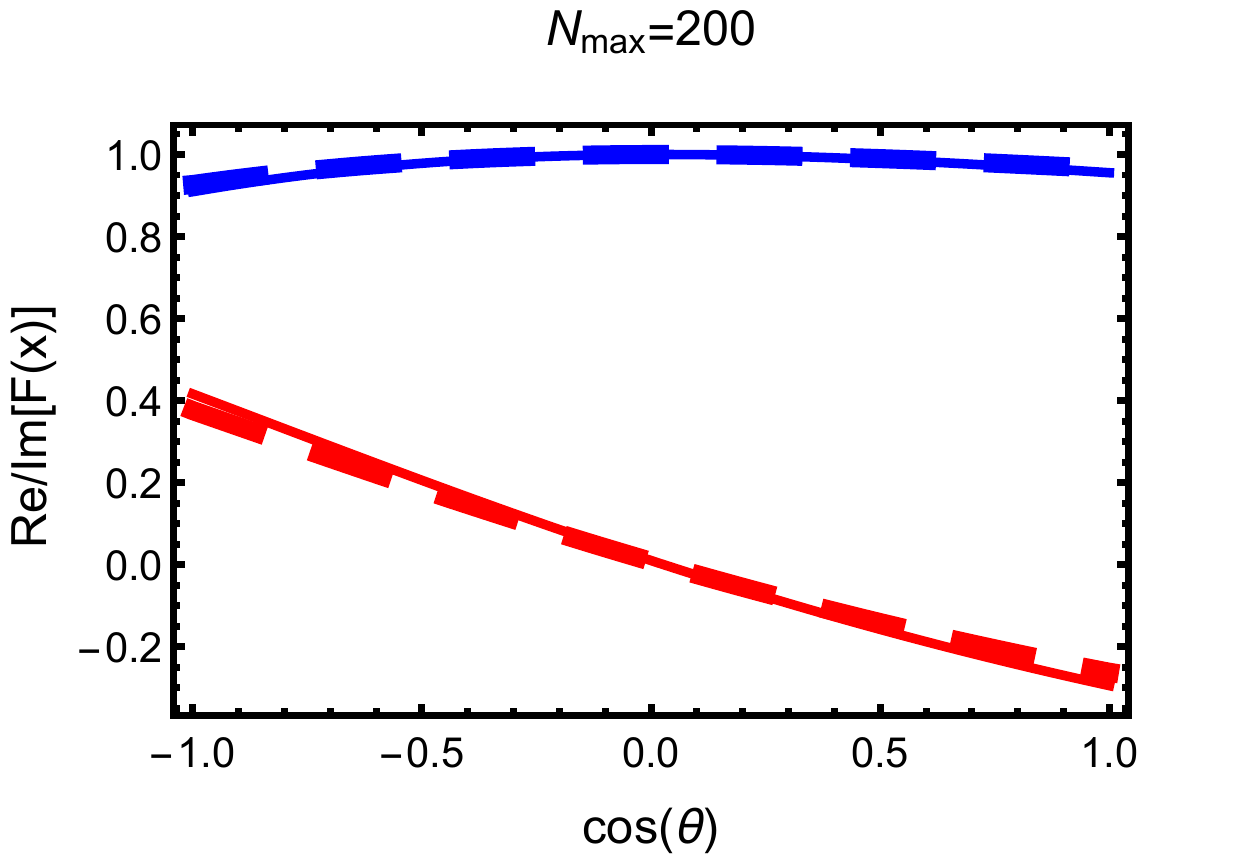}
\includegraphics[width=0.2075\textwidth]{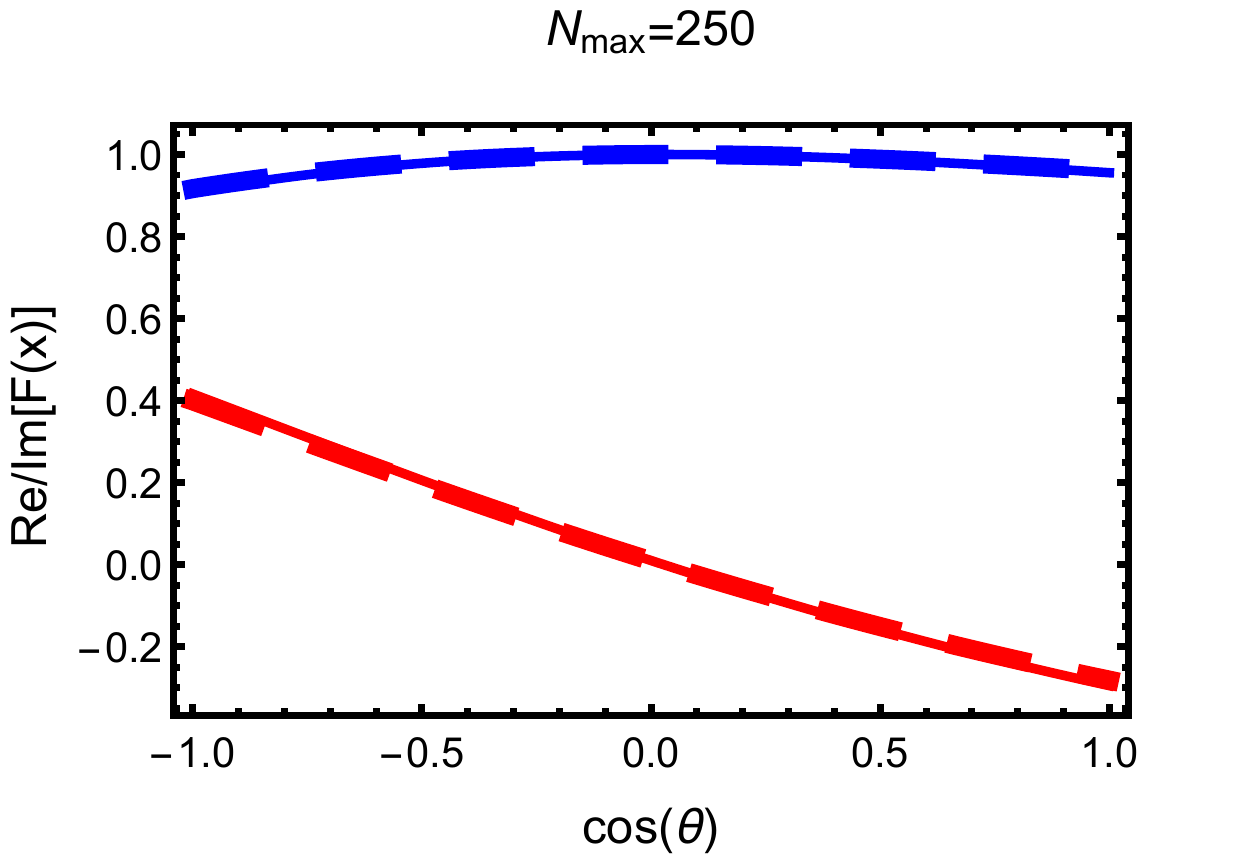}
\includegraphics[width=0.2075\textwidth]{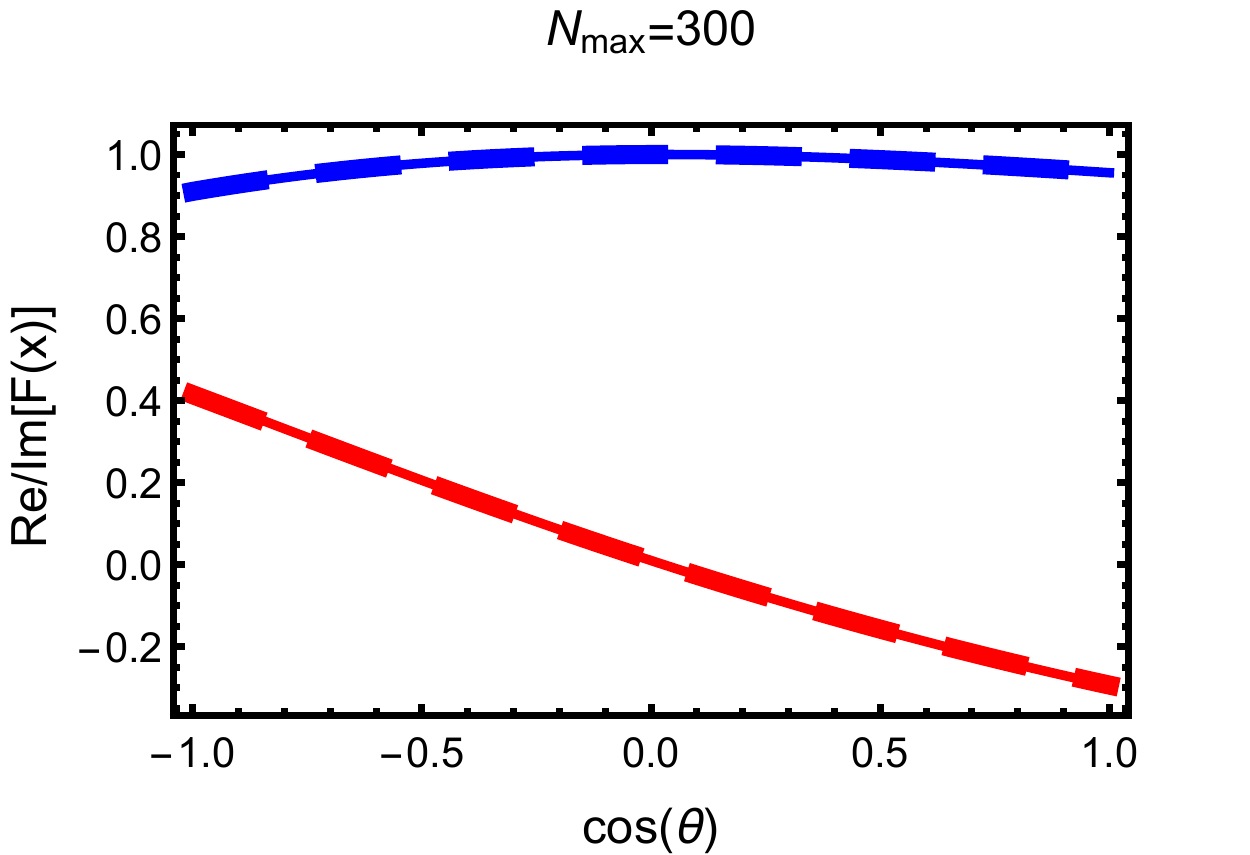}
\includegraphics[width=0.2075\textwidth]{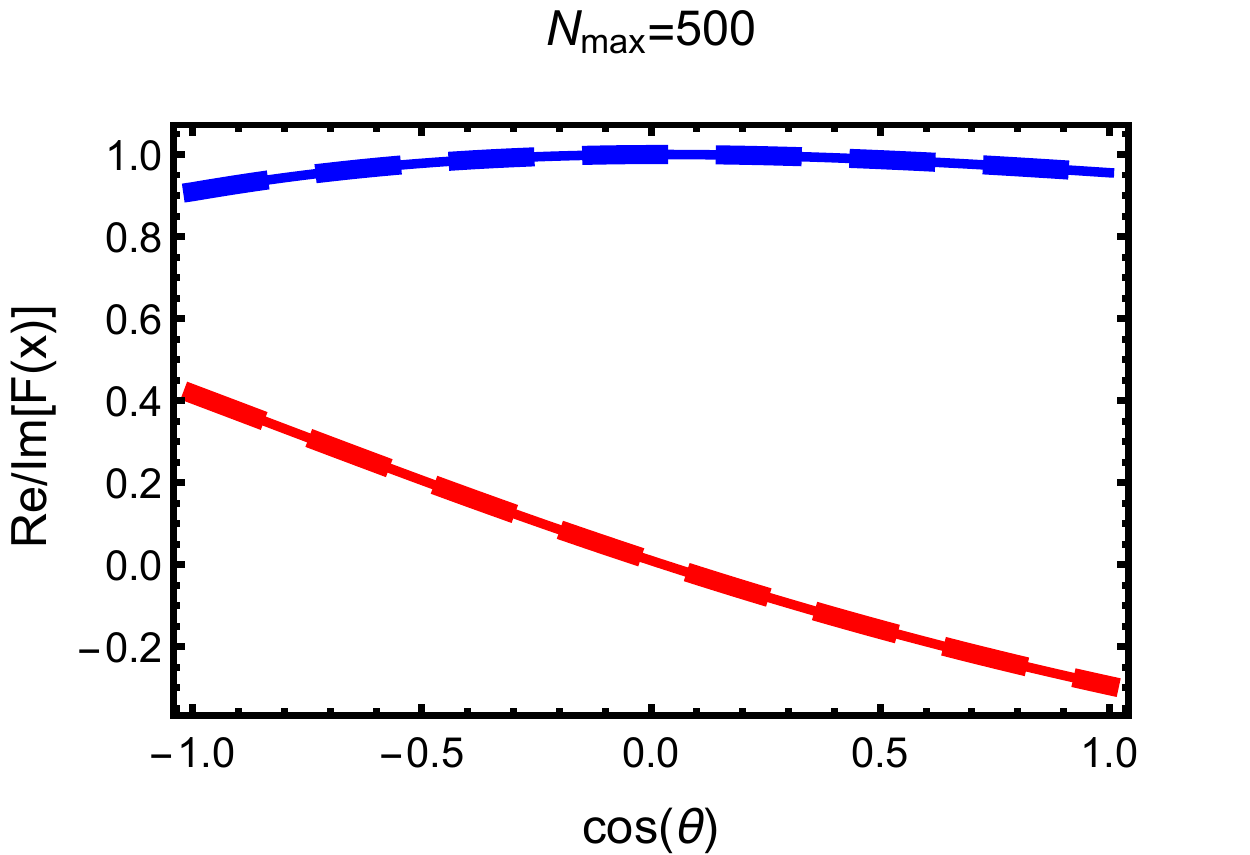} \\
% \caption{This is the continuation of Figure \ref{tab:FunctMinConvergencePlots1}. Convergence of the minimization of the functional (\ref{eq:FunctProblem}) is illustrated for the phases $e^{i \varphi_{2} (x)}$ and $e^{i \varphi_{3} (x)}$, which generate discrete symmetries for the toy-model (\ref{eq:DefLmax2ToyModel}).}
\captionof{figure}[fig]{These plots are the continuation of Figure \ref{tab:FunctMinConvergencePlots1}. The convergence of the numerical minimization of the functional (\ref{eq:FunctProblem}) is shown for the phases $e^{i \varphi_{2} (x)}$ and $e^{i \varphi_{3} (x)}$, which generate discrete ambiguities of the toy-model (\ref{eq:DefLmax2ToyModel}).}
\label{tab:FunctMinConvergencePlots2}
\end{table*}

\clearpage

\hspace*{-15pt} {\bf ACKNOWLEDGMENTS} \newline \newline
   The author (again, as in $2015$) wishes to thank the organizers for the hospitality, as well as for providing 
   a very relaxed and friendly atmosphere during the workshop. \newline
   This particular Bled-workshop takes a special place in this author's biography, since after $4$ months of battle with a very 
   bad knee-injury, the par\-ti\-ci\-pa\-tion in the workshop marked one of the first careful steps back into the world. Furthermore, the
   wonderful nature and environment of Bled itself turned out to be instrumental on the way of healing. By making the room 
   on the ground floor of the Villa Plemelj available, the organizers have provided a key to make par\-ti\-ci\-pa\-tion possible at all, and the 
   author wishes to express deep gratitude for that. The author's wife also wishes to thank the organizers for the possibility to stay 
   in Bled, as well as the nice hikes she made with the other participant's spouses. In fact, one early morning she was very brave and
   made a balloon ride over the lake of Bled. This author decided to include one of her aerial photographies into the proceeding. \newline
   This work was supported by the \textit{Deutsche Forschungsgemeinschaft} within the \\ SFB/TR16. \newline \newline
   %\vspace*{40pt}
   \hspace*{25pt}{\centering \includegraphics[width=0.90\textwidth]{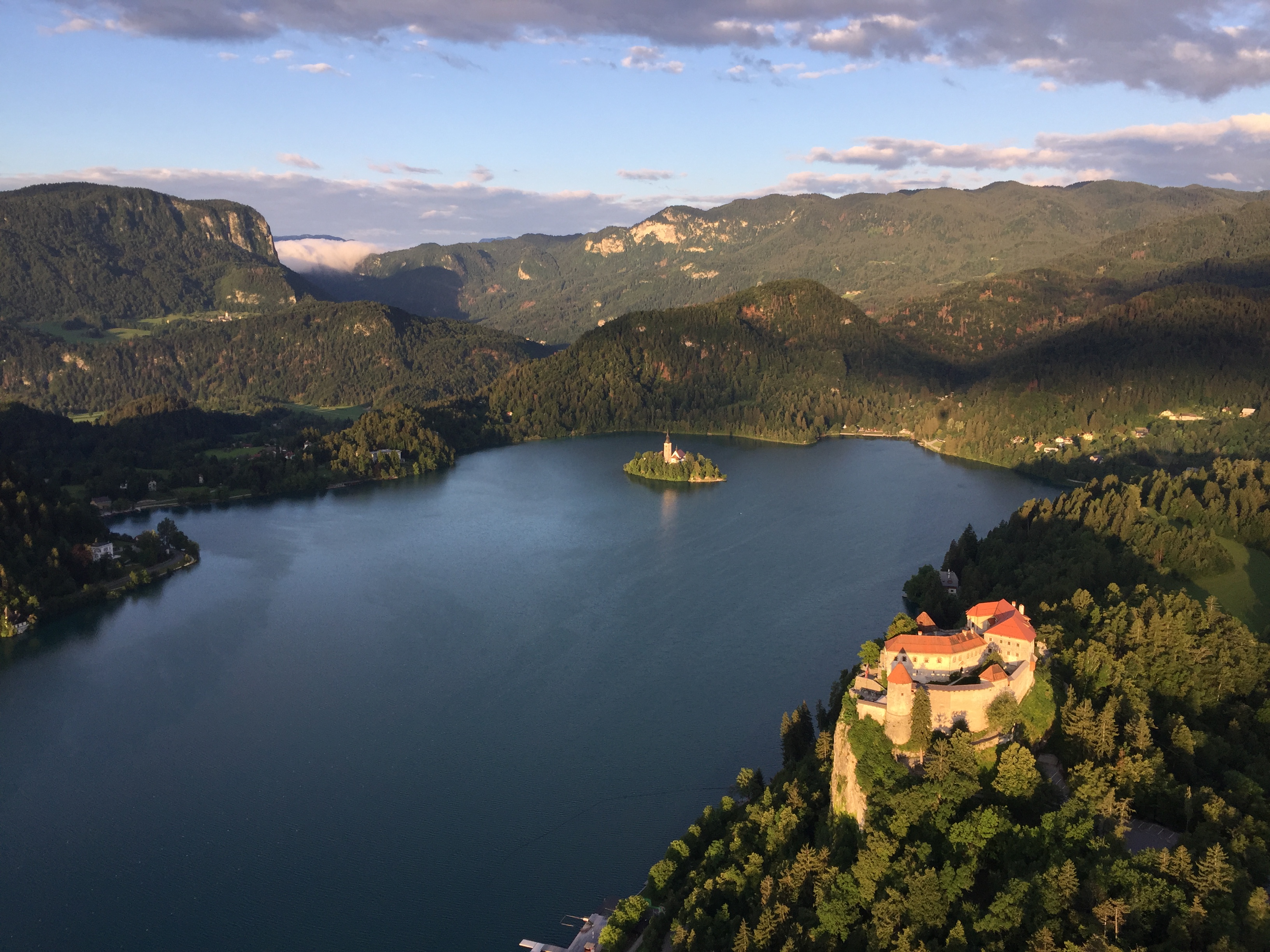}}

\newpage

\selectlanguage{english}

% \begin{table}[h]
% \centering
% \begin{tabular}{cc|cccccc||cc|cccccccc}
% \hline
% \hline
%   &  &  &  &  &  &  &  &  &  &  &  &  &  &  &  &  \\
% Type & $ \check{\Omega}^{\alpha} $ &  & $ \beta_{\alpha} $ &  & $ \gamma_{\alpha} $ &  & $ \delta_{\alpha} $ & Type & $ \check{\Omega}^{\alpha} $ &  & $ \beta_{\alpha} $ &  & $ \gamma_{\alpha} $ &  & $ \delta_{\alpha} $ &   \\
% \hline
%   &  &  &  &  &  &  &  &  &  &  &  &  &  &  &  &  \\
%   & $ I \left( \theta \right) $ &  & $ 0 $ &  & $ +1 $ &  & $ -2 $ &  & $ \check{O}_{x'} $ &  & $ 1 $ &  & $ +1 $ &  & $ -1 $ &   \\
%  S & $ \check{\Sigma} $ &  & $ 2 $ &  & $ -1 $ &  & $ -2 $ & BR & $ \check{O}_{z'} $ &  & $ 2 $ &  & $ 0 $ &  & $ -1 $ &   \\
%   & $ \check{T} $ &  & $ 1 $ &  & $ 0 $ &  & $ -1 $ &  & $ \check{C}_{x'} $ &  & $ 1 $ &  & $ +1 $ &  & $ -1 $ &   \\
%   & $ \check{P} $ &  & $ 1 $ &  & $ 0 $ &  & $ -1 $ &  & $ \check{C}_{z'} $ &  & $ 0 $ &  & $ +2 $ &  & $ -1 $ &   \\
% \hline
%   &  &  &  &  &  &  &  &  &  &  &  &  &  &  &  &  \\
%   & $ \check{G} $ &  & $ 2 $ &  & $ -1 $ &  & $ -1 $ &  & $ \check{T}_{x'} $ &  & $ 2 $ &  & $ 0 $ &  & $ -2 $ &   \\
%  BT & $ \check{H} $ &  & $ 1 $ &  & $ 0 $ &  & $ -1 $ & TR & $ \check{T}_{z'} $ &  & $ 1 $ &  & $ +1 $ &  & $ -2 $ &   \\
%   & $ \check{E} $ &  & $ 0 $ &  & $ +1 $ &  & $ -1 $ &  & $ \check{L}_{x'} $ &  & $ 1 $ &  & $ +1 $ &  & $ -2 $ &   \\
%   & $ \check{F} $ &  & $ 1 $ &  & $ 0 $ &  & $ -1 $ &  & $ \check{L}_{z'} $ &  & $ 0 $ &  & $ +2 $ &  & $ -2 $ &   \\
% \hline
% \hline
% \end{tabular}
% \caption{fhjnhfg}
% \label{tab:AngularDistributions}
% \end{table}

\end{document}